\documentclass[12pt]{article}

\usepackage{latexsym}
\usepackage{epsfig,amssymb,euscript}
\usepackage{amsmath}
\usepackage{mathrsfs}

\numberwithin{equation}{section}

\def\be{\begin{equation}}
\def\ee{\end{equation}}
\def\bea{\begin{eqnarray}}
\def\eea{\end{eqnarray}}

\newcommand\Q{\mathcal{Q}}
\newcommand\R{\mathbb{R}}
\newcommand\Z{\mathbb{Z}}

\newcommand\C{\mathbb{C}}

\newcommand\diff{\mathrm{d}}

\newcommand{\de}{\partial}
\newcommand{\vol}{\mathrm{vol}}
\newcommand{\nn}{\nonumber \\}
\newcommand{\reef}[1]{(\ref{#1})}

\newcommand{\dd}{\diff}
\newcommand{\me}{\mathrm{e}}
\newcommand{\ii}{\mathrm{i}}
\newcommand{\del}{\partial}
\newcommand{\delb}{\bar{\partial}}
\newcommand{\bbZ}{\mathbb{Z}}
\newcommand{\bbR}{\mathbb{R}}
\newcommand{\bbC}{\mathbb{C}}

\newlength{\sswidth}
\newcommand{\sla}[1]{
   \settowidth{\sswidth}{$#1$}
   \mbox{$\not{\hspace*{-0.3\sswidth}#1}$}}

\DeclareMathOperator{\im}{Im}
\DeclareMathOperator{\re}{Re}
\DeclareMathOperator{\SU}{\mathit{SU}}

\DeclareMathOperator{\GL}{\mathit{GL}}

\DeclareMathOperator{\Spin}{\mathit{Spin}}

\newcommand{\id}{1}

\newcommand{\Lgen}{\mathbb{L}}

\DeclareMathOperator{\Cliff}{Cliff}

\newcommand{\Bc}{B}
\newcommand{\Gstar}{\star_G}
\newcommand{\SC}[1]{\Gamma(#1)}
\newcommand{\tpsi}{\tilde{\psi}}
\newcommand{\Ered}{E^{\text{red}}}
\newcommand{\Mred}{M^{\text{red}}}
\newcommand{\gred}{g^{\text{red}}}
\newcommand{\LJ}{\tilde{\mathcal{J}}_1}
\newcommand{\RJ}{\tilde{\mathcal{J}}_2}
\newcommand{\Lred}{\tilde{\Omega}_1}
\newcommand{\Rred}{\tilde{\Omega}_2}
\newcommand{\etar}{\tilde{\eta}}

\newcommand{\mukai}[2]{\langle{#1},{#2}\rangle}
\newcommand{\FF}{\Omega_-}
\newcommand{\Btheta}{\bar{\theta}}
\newcommand{\Bphi}{\bar{\phi}}

\begin{document}

\begin{titlepage}

\begin{flushright}
Imperial-TP-2009-JG-05
\end{flushright}

\begin{center}
\today\vskip 1.5cm
{\Large\bf $AdS_5$ Solutions of Type IIB Supergravity \\*[2mm]
  and Generalized Complex Geometry}
\\
\vskip 1cm
{Maxime Gabella$^1$, ~ Jerome P. Gauntlett$^2$, ~ Eran Palti$^1$, }\\[2mm]
{James Sparks$^{3}$ ~and~ Daniel Waldram$^2$}\\
\vskip 1cm
1: {\em Rudolf Peierls Centre for Theoretical Physics,\\
 University of Oxford, \\
1 Keble Road, Oxford OX1 3NP, U.K.}\\
\vskip 0.5cm
2: {\em Theoretical Physics Group, Blackett Laboratory,\\
Imperial College, London SW7 2AZ, U.K.}\\
\vskip 0.2cm
{\em The Institute for Mathematical Sciences,\\
Imperial College, London SW7 2PE, U.K.}\\
\vskip 0.5cm
3: {\em Mathematical Institute, University of Oxford,\\
 24-29 St Giles', Oxford OX1 3LB, U.K.}\\
%\vskip 1.4cm
\end{center}
\vskip 0.5cm

\begin{abstract}
\vskip 0.4cm
We 
use the formalism of generalized geometry to study the generic
supersymmetric $AdS_5$ solutions of type IIB supergravity that are
dual to $\mathcal{N}=1$ superconformal field theories (SCFTs) in
$d=4$. Such solutions have an associated six-dimensional  
generalized complex cone geometry that is an extension of
Calabi-Yau cone geometry.  
We identify generalized vector fields dual to the dilatation and
$R$-symmetry of the dual SCFT and show that they are generalized
holomorphic on the cone. We carry out a generalized reduction of the
cone to a transverse four-dimensional space and show that this is also
a generalized complex geometry, which is an extension of
K\"ahler-Einstein geometry. 
Remarkably, provided the five-form flux is non-vanishing, the cone is
symplectic. The symplectic structure can be used to obtain Duistermaat-Heckman
type integrals for the central charge of the dual SCFT and the
conformal dimensions of operators dual to BPS wrapped D3-branes. We
illustrate these results using the Pilch-Warner solution. 
\end{abstract}
\vfill
\end{titlepage}
\pagestyle{plain}
\setcounter{page}{1}
\newcounter{bean}
\baselineskip18pt

%%%%%%%%%%%%%%%%%%%%%%%%%%%%%%%%%%%%%%%%%%%%%%%%%%%%%
\section{Introduction}
%%%%%%%%%%%%%%%%%%%%%%%%%%%%%%%%%%%%%%%%%%%%%%%%%%%%%

Supersymmetric $AdS_5\times Y$ solutions of type IIB supergravity,
where $Y$ is a compact Riemannian manifold, are dual to supersymmetric
conformal field theories (SCFTs) in $d=4$ spacetime dimensions with
(at least) ${\cal N}=1$ supersymmetry. An important special subclass
is when $Y$ is a five-dimensional Sasaki-Einstein manifold $SE_5$, and
the only non-trivial flux is the self-dual five-form. Recall that, by
definition, the six-dimensional cone metric with  
base given by the $SE_5$ space is a Calabi-Yau cone, and that the dual
SCFT arises from D3-branes located at the apex of this cone. 
There has been much progress in understanding the AdS/CFT
correspondence in this setting. For example, there are rich sets of
explicit $SE_5$ metrics \cite{Gauntlett:2004zh}-\cite{Cvetic:2005ft},
and there are also powerful constructions using toric
geometry. Moreover, for the toric case, the corresponding dual SCFTs
have been identified, {\it e.g.}
\cite{Benvenuti:2004dy}-\cite{Franco:2005sm}. 

A key aspect of this progress has been the appreciation that the abelian $R$-symmetry, which all ${\cal N}=1$ SCFTs in $d=4$
possess, contains important information about the SCFT. For example, the $a$ central charge is fixed by the $R$-symmetry, 
as are the anomalous dimensions of (anti-)chiral primary operators \cite{Anselmi:1997ys}. It is also known that the $R$-symmetry
can be identified via the procedure of $a$-maximization, which, roughly, says that the correct $R$-symmetry is
the one that maximizes the value of $a$ over all possible admissible $R$-symmetries \cite{kenbrian}.
For the solutions of type IIB supergravity with $Y= SE_5$, the $R$-symmetry manifests itself as a canonical Killing vector $\xi$ on
$SE_5$. This defines a Killing vector on the Calabi-Yau cone, also denoted by $\xi$, which is a real
holomorphic vector field. The Calabi-Yau cone is K\"ahler, and hence symplectic, and $Y$ admits a corresponding contact structure for
which $\xi$ is the Reeb vector.
When $Y=SE_5$ the $a$ central charge is inversely proportional to the volume
of $SE_5$, and in \cite{Martelli:2005tp, Martelli:2006yb} several geometric formulae for $a$ in terms of 
$\xi$ were derived. Analogous geometric formulae for the conformal
dimension of the chiral primary operator 
dual to a D3-brane wrapped 
on a supersymmetric submanifold $\Sigma_3\subset Y$ were
also presented.
Of particular interest here are the formulae 
that show that using symplectic geometry these quantities can be written as Duistermaat-Heckman 
integrals on the cone and hence can be evaluated by localization.
In addition to providing a geometrical interpretation of a-maximization, these formulae and others in
\cite{Martelli:2005tp, Martelli:2006yb} also provide practical methods for calculating 
quantities of physical interest without needing the full explicit Sasaki-Einstein metric
(which, apart from some special classes of solution, remains out of reach).

The focus of this paper is on $AdS_5\times Y$ solutions with $Y$ more general than $SE_5$. Most known 
solutions are actually part of continuous families of solutions containing a Sasaki-Einstein
solution and correspond to exactly marginal deformations of the corresponding SCFT.
For example, starting with a toric $SE_5$ solution one can construct new $\beta$-deformed solutions 
using the techniques of \cite{Lunin:2005jy}. There is also the ``Pilch-Warner solution'' explicitly 
constructed in \cite{Pilch:2000ej} (based on \cite{Khavaev:1998fb}). It has been shown numerically 
in \cite{Halmagyi:2005pn} that the $\bbZ_2$ orbifold of the Pilch-Warner solution 
is part of a continuous family of solutions that includes the Sasaki-Einstein $AdS_5\times T^{1,1}$ solution.
Using the results of \cite{Leigh:1995ep,Benvenuti:2005wi} this should be part of a larger family of continuous solutions that are
yet to be found. Similarly, in addition to the $\beta$-deformations of the 
$AdS_5\times S^5$ solution there are additional deformations \cite{Leigh:1995ep}
that are also not yet constructed (a perturbative analysis 
was studied in \cite{Aharony:2002hx}). Having a better understanding
of the geometry underlying general $AdS_5\times Y$ solutions could be
useful for finding these deformed solutions but more generally could
be useful in constructing new solutions that are not connected with
Sasaki-Einstein geometry at all. 

The first detailed analysis of supersymmetric $AdS_5\times Y$
solutions of type IIB supergravity, for general $Y$ with all fluxes
activated, was carried out in \cite{Gauntlett:2005ww}. The conditions
for supersymmetry boil down to a set of Killing spinor equations on
$Y$ for two spinors (when $Y=SE_5$ there is only one such spinor). By
analysing these equations a set of necessary and sufficient conditions
for supersymmetry were established. In light of the progress
summarized above for the Sasaki-Einstein case, it is natural to
investigate the associated geometry of the cone over $Y$, and that is
the principal purpose of this paper. 

As we shall discuss in detail, the cone $X$ admits a specific kind of
generalized complex geometry.  
Aspects of this geometry, restricting to a class of
$\SU(2)$-structures, were first studied
in~\cite{Minasian:2006hv,Butti:2007aq}. By viewing $AdS_5\times Y$ as
a supersymmetric warped product $\bbR^{3,1}\times X$, one sees
immediately that the cone admits two compatible generalized almost
complex structures~\cite{Grana:2004bg,Grana:2005sn}, or equivalently
two  compatible pure spinors, $\Omega_\pm$. In fact $\dd \Omega_-=0$, so
that $\Omega_-$ defines an integrable generalized complex geometry,
while $\dd\Omega_+$ is related to the RR flux. The cone is thus
generalized Hermitian, and it is also generalized Calabi-Yau in the
sense of~\cite{hitchin}.   

Here, we will identify a generalized vector field $\xi$ on the cone
that is dual to the $R$-symmetry and another that is dual to the
dilatation symmetry of the dual SCFT and show that they are both
generalized holomorphic vector fields on the cone  (with respect to
the integrable generalized complex structure). This precisely
generalizes known results for the Sasaki-Einstein case. 
We also note that all supersymmetric $AdS_5\times Y$ solutions satisfy
the condition of~\cite{Minasian:2006hv} that there is an
$\SU(2)$-structure on the cone. 

In the Sasaki-Einstein case, one can carry out a symplectic reduction
of the Calabi-Yau cone to obtain a four-dimensional transverse
K\"ahler-Einstein space which, in general, is only locally
defined. Constructing locally defined K\"ahler-Einstein spaces has
been a profitable way to construct Sasaki-Einstein manifolds, {\it
  e.g.} \cite{Gauntlett:2004hh}. Here we will show, 
using the formalism of~\cite{bcg1,bcg}, that for general
$Y$ there is an analogous reduction of the corresponding
six-dimensional generalized Calabi-Yau cone geometry to a
four-dimensional space, which again is only locally defined in
general, that is generalized Hermitian. More precisely, the
four-dimensional geometry admits two compatible generalized almost
complex structures, one of which is integrable. 

We present explicit expressions for the pure spinors $\Omega_\pm$ associated with the six-dimensional cone
in terms of the Killing spinor bilinears presented
in \cite{Gauntlett:2005ww}. We shall comment upon how $\Omega_-$, associated with the integrable generalized complex structure, 
contains information on the mesonic moduli space of the dual SCFT and also, briefly, on some
relations connected with generalized holomorphic objects and dual BPS operators.

By analysing the pure spinor $\Omega_+$, associated with the non-integrable complex structure,
and focusing on the case when the five-form flux is non-vanishing, we show that, perhaps somewhat surprisingly, the cone
is symplectic. We shall see that $Y$ is a contact manifold and that the vector part, $\xi_v$, of the
generalized vector $\xi$, which also defines a Killing vector on $Y$, is the Reeb vector field associated
with the contact structure. We show that the symplectic structure 
can be used to obtain Duistermaat-Heckman type integrals for the central charge $a$
of the dual SCFT and also for the conformal dimensions of operators dual to wrapped BPS D3-branes.
Once again these results precisely generalize those for the Sasaki-Einstein case. Some of these results were first presented in \cite{shortpaper}; 
here we will provide additional details and 
also show how they are related to the generalized geometry on the cone.

Finally, we will illustrate some of our results 
using the Pilch-Warner solution. 
The paper begins with a review of generalized geometry and it ends with three appendices containing some details about our conventions,
some technical derivations, and a brief discussion of the Sasaki-Einstein case.

%%%%%%%%%%%%%%%%%%%%%%%%%%%%%%%%%%%%%%%%%%%%%%%%%%%%%
\section{Generalized geometry}
%%%%%%%%%%%%%%%%%%%%%%%%%%%%%%%%%%%%%%%%%%%%%%%%%%%%%
We begin by reviewing some aspects of generalized
complex geometry \cite{hitchin}, to fix conventions and notation.
For further details see, for example, \cite{gualtieri}.

%%%%%%%%%%%%%%%%%%%%%%%%%%%%%%%%%%%%%%%%%%%%%%%%%%%%%
\subsection{The generalized tangent and spinor bundles}\label{sec:gentangent}
%%%%%%%%%%%%%%%%%%%%%%%%%%%%%%%%%%%%%%%%%%%%%%%%%%%%%
Generalized geometry starts with the generalized tangent bundle $E$
over a manifold $X$, which is a particular extension of
$TX$ by $T^*X$ obtained by twisting with a \emph{gerbe}. A gerbe is
simply a higher degree version of a $U(1)$ bundle with unitary
connection. Just as topologically a $U(1)$ bundle is determined
by its first Chern class, the topology of a gerbe is determined by a class
in $H^3(X,\bbZ)$.
To define a gerbe~\cite{Hitchin:2005in}, one begins with an open cover
$\{U_i\}$ of $X$ together with a set of functions $g_{ijk}:U_i\cap
U_j\cap U_j\rightarrow U(1)$ defined on triple overlaps. These are
required to satisfy $g_{ijk}={g_{jik}}^{-1}={g_{ikj}}^{-1}={g_{kji}}^{-1}$,
together with the cocycle condition
$g_{jkl}{g_{ikl}}^{-1}{g_{ijl}}{g_{ijk}}^{-1}=1$ on quadruple overlaps. A
\emph{connective structure} \cite{Hitchin:2005in} on a gerbe is a
collection of one-forms $\Lambda_{(ij)}$ defined on double overlaps
$U_i\cap U_j$ satisfying 
$\Lambda_{(ij)}+\Lambda_{(jk)}+\Lambda_{(ki)}=-(2\pi \ii l_s^2)
g_{ijk}^{-1}\diff g_{ijk}$ on triple overlaps. In turn, a
\emph{curving} is a collection of two-forms $\Bc_{(i)}$ on $U_i$
satisfying 
\bea
\label{splitting}
\Bc_{(j)} - \Bc_{(i)} = \diff \Lambda_{(ij)}~.
\eea
It follows that $\diff \Bc_{(j)}=\diff \Bc_{(i)}=H$
is a closed global three-form on $X$, called the
\emph{curvature}, and, in cohomology, $\frac{1}{(2\pi l_s)^2}H\in
H^3(X,\bbZ)$ (in the normalization that we shall use in this paper).
In string theory, the collection of two-forms
$\Bc_{(i)}$, which  we write simply as $\Bc$, is the NS $B$-field and
$H$ is its curvature.

The generalized tangent bundle $E$ is an extension of $TX$ by $T^*X$
\begin{equation}
\label{eq:Edef}
   0 \longrightarrow T^*X \longrightarrow E
      \stackrel{\pi}{\longrightarrow} TX \longrightarrow 0~.
\end{equation}
Locally, sections of $E$, which we refer to as generalized tangent vectors, may be written as $V=x+\lambda$, where
$x\in\SC{TX}$ and $\lambda\in\SC{T^*X}$. More precisely, in going from one
coordinate patch $U_i$ to another $U_j$ the extension is defined by
the connective structure
\begin{equation}
\label{eq:S-patch}
   x_{(i)} + \lambda_{(i)}
      = x_{(j)} + \big( \lambda_{(j)} - i_{x_{(j)}}\dd\Lambda_{(ij)} \big)~.
\end{equation}
The bundle $E$ is in fact isomorphic to $TX\oplus T^*X$. However, the
isomorphism is not canonical but depends on a choice of splitting, defined by a two-form curving $\Bc$ satisfying
(\ref{splitting}). It follows that
\begin{equation}
   x+(\lambda-i_x\Bc) \in \SC{TX\oplus T^*X} ~.
\end{equation}
Thus the definition~\eqref{eq:S-patch} of $E$ can be
viewed as encoding the patching of a class of two-form curvings $\Bc$.

Writing $d=\dim_\R X$, there is a natural $O(d,d)$-invariant metric
$\mukai{\cdot}{\cdot}$ on $E$, given by
\begin{equation}
\label{eq:eta}
%   \eta(V,W) \equiv 
   \mukai{V}{W}= \tfrac{1}{2}(i_x\mu+i_y\lambda)~,
\end{equation}
where $V=x+\lambda$, $W=y+\mu$, or in two-component notation, 
% $\eta(V,W)=V^T\eta W$ with
%
\begin{equation}
   \mukai{V}{W} = \big( x \quad \lambda \big)
         \begin{pmatrix} 0 & \frac{1}{2} \\ \frac{1}{2} &
            0 \end{pmatrix}
         \begin{pmatrix} y \\ \mu \end{pmatrix}~.
%   \eta = \frac{1}{2}\begin{pmatrix} 0 & 1 \\ 1 & 0 \end{pmatrix}~.
\end{equation}
This metric is invariant under $O(d,d)$ transformations acting on the
fibres of $E$, defining a canonical $O(d,d)$-structure.
A general element $O\in O(d,d)$ may be
written in terms of  $d \times d$ matrices $a$, $b$, $c$, and $d$ as
\begin{equation}
\label{eq:oddel}
   O = \begin{pmatrix} a & b \\ c & d \end{pmatrix} \ ,
\end{equation}
under which a general element $V \in E$ transforms by
\begin{equation}
\label{eq:Odd}
\begin{aligned}
   V = \begin{pmatrix} x \\ \lambda \end{pmatrix}
      &\mapsto OV = \begin{pmatrix} a & b \\ c & d \end{pmatrix}
         \begin{pmatrix} x \\ \lambda \end{pmatrix}~.
\end{aligned}
\end{equation}
The requirement that $\mukai{OV}{OV}=\mukai{V}{V}$ implies $a^Tc+c^Ta=0$,
$b^T d + d^T b=0$ and $a^T d+c^T b=1$. Note that the $GL(d)$ action on
the fibres of $TX$ and $T^*X$ embeds as a subgroup of
$O(d,d)$. Concretely it maps
\begin{equation}
\label{eq:diffeo}
   V \mapsto V' =
      \begin{pmatrix} a & 0 \\ 0 & a^{-T} \end{pmatrix}
      \begin{pmatrix} x \\ \lambda \end{pmatrix}~,
\end{equation}
where $a\in\GL(d)$. Given a two-form $\omega$, one also has the
abelian subgroup
\begin{equation}
\label{eq:Btransform}
   \me^\omega =
      \begin{pmatrix} \id & 0 \\ \omega & \id \end{pmatrix}
   \qquad \text{such that} \qquad
   V = x + \lambda \mapsto V' = x + (\lambda-i_x\omega)~.
\end{equation}
This is usually referred to as a $B$-transform. Given a bivector
$\beta$ one can similarly define another abelian subgroup of
$\beta$-transforms
\begin{equation}
\label{eq:betatransform}
   \me^\beta =
      \begin{pmatrix} \id & \beta \\ 0 & \id \end{pmatrix}
   \qquad \text{such that} \qquad
   V = x + \lambda \mapsto V' = (x + i_\lambda\beta)+  \lambda~.
\end{equation}
Note that the patching~\eqref{eq:S-patch} corresponds to a $B$-transform
with $\omega=\dd\Lambda_{(ij)}$. Similarly, the splitting isomorphism
between $E$ and $TX\oplus T^*X$ defined by $\Bc$ is also a
$B$-transform
\begin{equation}
\label{eq:split-iso}
   E \underset{\me^{-\Bc}}{\stackrel{\me^\Bc}{\rightleftarrows}}
      TX\oplus T^*X~.
\end{equation}

There is a natural bracket on generalized vectors known as the
Courant bracket, which encodes the differentiable structure of $E$. It
is defined as
\begin{equation}
\label{eq:Courant}
   [V,W]= [ x + \lambda, y + \mu ]
     = [x,y]_{\rm Lie} + \mathcal{L}_x\mu - \mathcal{L}_y\lambda
        - \tfrac{1}{2} \dd\left(i_x\mu - i_y\lambda\right)~,
\end{equation}
where $[x,y]_{\rm Lie}$ is the usual Lie bracket between vectors and
$\mathcal{L}_x$ is the Lie derivative along $x$. The Courant bracket
is invariant under the action of diffeomorphisms and $B$-shifts
$\omega$ that are closed, $\dd\omega=0$, giving an automorphism
group which is a semi-direct product $\mathrm{Diff}(X)\ltimes
\Omega^2_{\mathrm{closed}}(X)$. Note, however, that in string theory
only $B$-shifts by the curvature of a unitary line bundle
on $X$ are gauge symmetries, as opposed to shifts by arbitrary closed
two-forms, leading to a smaller automorphism
group. Under an infinitesimal
diffeomorphism generated by a vector field $x$ and a $B$-shift with
$\omega=\dd\lambda$, one has the generalized Lie derivative by
$V=x+\lambda$ on a generalized vector field $W=y+\mu$
\begin{equation}
\label{eq:Lgen}
   \delta W \equiv \Lgen_V W
      = [x,y]_{\rm Lie} + (\mathcal{L}_x\mu - i_y\dd\lambda)~.
\end{equation}
This is also known as the Dorfman bracket $[V,W]_D$, the
anti-symmetrization of which gives the Courant
bracket~\eqref{eq:Courant}. 
Note that since the metric $\mukai{\cdot}{\cdot}$ is invariant under
$O(d,d)$ transformations its generalized Lie derivative vanishes. 
Given a particular choice of splitting~\eqref{splitting} defined by
$\Bc$, the Courant bracket on $E$ defines a Courant bracket on
$TX\oplus T^*X$, known as the \emph{twisted} Courant
bracket. It is given by
\begin{equation}
\begin{aligned}
\label{eq:twistedCourant}
   [x+\lambda,y+\mu]_H
      &= \me^{\Bc}\,[\me^{-\Bc}(x+\lambda),\me^{-\Bc}(y+\mu)] \\
      &= [x+\lambda,y+\mu] + i_yi_x H~,
\end{aligned}
\end{equation}
where by an abuse of notation we are writing $x+\lambda$ and $y+\mu$ for sections
of $TX\oplus T^*X$ whereas above they were sections of $E$.

Given the metric $\mukai{\cdot}{\cdot}$, one can define $\Spin(d,d)$
spinors in the usual way. Since the volume element in $\Cliff(d,d)$
squares to one, one can define two helicity spin bundles $S_\pm(E)$ as
the $\pm1$ eigenspaces, and thus take spinors to be Majorana-Weyl. A
section of $S_\pm(E)$ on $U_i$ can be identified with a even- or
odd-degree polyform $\Omega_\pm\in\Omega^{\text{even/odd}}(X)$
restricted to $U_i$,  with the Clifford action of $V\in \SC E$ given by
\begin{equation}
\label{cliff}
   V\cdot \Omega_\pm
       = i_x \Omega_\pm
         + \lambda\wedge \Omega_\pm \, .
\end{equation}
It is easy to see that
\begin{equation}
   (V\cdot W+W\cdot V)\cdot\Omega_\pm = 2\mukai{V}{W}\,\Omega_\pm \, ,
\end{equation}
as required. Using this Clifford action the
$B$-transform~\eqref{eq:Btransform} on spinors is given by
\begin{equation}
   \Omega_\pm \mapsto \me^\omega \Omega_\pm \ ,
\end{equation}
where the exponentiated action is by wedge product. The
patching~\eqref{eq:S-patch} of $E$ then implies that
\begin{equation}
\label{spinorpatch}
   \Omega_\pm^{(i)}
      = \me^{\dd\Lambda_{(ij)}}\Omega_\pm^{(j)} \, .
\end{equation}
Furthermore a splitting $\Bc$ also induces an isomorphism between
$S_\pm(E)$ and $S_\pm(TX\oplus TX^*)$
\begin{equation}
   S_\pm(E) \underset{\me^{-\Bc}}{\stackrel{\me^\Bc}{\rightleftarrows}}
      S_\pm(TX\oplus T^*X)~,
\end{equation}
again by the action of the exponentiated wedge product. If $\Omega_\pm$
is a section of $S_\pm(E)$, we will sometimes write
$\Omega_\pm^\Bc\equiv \me^\Bc\Omega_\pm$ for the corresponding section of
$S_\pm(TX\oplus T^*X)$ defined by the splitting $\Bc$.
The real $\Spin(d,d)$-invariant spinor bilinear on sections of
$S_\pm(E)$ is a top form given by the \emph{Mukai pairing}
\begin{equation}
   \mukai{\Omega_\pm}{\Psi_\pm} \equiv
     \left(\Omega_\pm\wedge \lambda(\Psi_\pm)\right)_{\text{top}}~,
\end{equation}
where one defines the operator $\lambda$
\bea\label{eq:lambda}
\lambda(\Psi^\pm_m) \equiv (-1)^{\mathrm{Int}[m/2]}\Psi^\pm_m~,
\eea
with $\Psi_m$ the degree $m$ form in $\Psi_\pm$.
The Mukai paring is invariant under $B$-transforms:
$\mukai{\me^\omega\Omega_\pm}{\me^\omega\Psi_\pm} =\mukai{\Omega_\pm}{\Psi_\pm}$.
For $d=6$ the
bilinear is anti-symmetric.
The usual action of the exterior derivative on the component forms of
$\Omega_\pm$ is compatible with the patching~\eqref{spinorpatch} and
defines an action
\begin{equation}
\label{eq:dd-Dirac}
   \dd:S_\pm(E)\to S_\mp(E)~,
\end{equation}
while the generalized Lie derivative on spinors is given by
\begin{equation}
\label{eq:genLS}
   \Lgen_V \Omega_\pm = \mathcal{L}_x \Omega_\pm + \dd\lambda\wedge\Omega_\pm
       = \dd (V\cdot\Omega_\pm) + V\cdot \dd\Omega_\pm~.
\end{equation}
Note that given a splitting $\Bc$ the operator 
on $\Omega^B_\pm\in S_\pm(TX\oplus T^*X)$ corresponding to $\dd$ is
$\dd_H$ defined by 
\begin{equation}
\label{eq:diffH}
   \dd_H \Omega^B_\pm \equiv \me^{\Bc}\dd(\me^{-\Bc}\Omega^B_\pm)
      = \left(\dd - H\wedge{}\right)\Omega^B_\pm~,
\end{equation}
where $H=\dd\Bc$. Furthermore one has 
\be
\Lgen_V\Omega=\me^{-B}\left(\Lgen_{V^B}-i_xH\wedge{}\right)\Omega^B~,
\ee
where $V^B=\me^BV=x+(\lambda-i_xB)$.

Finally, we note that there is actually a slight subtlety in the
relation between generalized spinors and polyforms. Given the
embedding~\eqref{eq:diffeo} in $O(d,d)$ of the $\GL(d)$ action on the
fibres of $TX$ one actually finds that the Clifford
action~\eqref{cliff} implies that on $U_i$ we can identify $S_\pm(E)$
with
$|\Lambda^dT^*X|^{-1/2}\otimes\Lambda^{\text{even/odd}}T^*X$; that is,
there is an additional factor of the determinant bundle
$|\Lambda^dT^*X|$. (This factor is the source, for instance, of the
fact that the Mukai pairing is a top form, rather than a scalar.) This
bundle is trivial, so generalized spinors can indeed be written as
polyforms patched by~\eqref{spinorpatch}, but there is no natural
isomorphism to make this identification. The simplest solution,
and one which will also allow us to incorporate the dilaton in a
natural way, is to extend the $O(d,d)$ action to a conformal action
$O(d,d)\times\bbR^+$. One can then define a family of spinor bundles
$S^k_\pm(E)$ transforming with weight $k$ under the conformal factor
$\bbR^+$; that is, with sections transforming as
$\Omega_\pm\to\rho^k\Omega_\pm$ where $\rho\in\bbR^+$. If one
embeds the $\GL(d)$ action on $TX$ in $O(d,d)$ as in~\eqref{eq:diffeo}
and, in addition, makes a conformal scaling by $\rho=\det a$ then
sections of $S^{-1/2}_\pm(E)$ can be directly identified with
polyforms patched by~\eqref{spinorpatch}.

%%%%%%%%%%%%%%%%%%%%%%%%%%%%%%%%%%%%%%%%%%%%%%%%%%%%%
\subsection{Generalized metrics and complex structures}
\label{sec:genmetric}
%%%%%%%%%%%%%%%%%%%%%%%%%%%%%%%%%%%%%%%%%%%%%%%%%%%%%

A \emph{generalized metric} $G$ on $E$ is the generalized geometrical
equivalent of a Riemannian metric on $TX$. We have seen that there
is a natural $O(d,d)$ structure on $E$ defined by the metric
$\mukai{\cdot}{\cdot}$~\eqref{eq:eta}. The generalized metric $G$
defines an $O(d)\times O(d)$ substructure. It splits $E=C_+\oplus C_-$
such that the metric $\mukai{\cdot}{\cdot}$ gives a positive-definite
metric on $C_+$ and a negative-definite metric on $C_-$, corresponding
to the two $O(d)$ structure groups. One can define $G$ as a product
structure on $E$; that is, $G:E\to E$ with $G^2=\id$ and
$\mukai{GU}{GV}=\mukai{U}{V}$, so that $\frac{1}{2}(\id\pm G)$ project
onto $C_\pm$. In general $G$ has the form
\begin{equation}
\label{eq:genmetric}
   G = \begin{pmatrix}
         g^{-1} B & g^{-1} \\
         g - B g^{-1} B & - B g^{-1}
      \end{pmatrix}
     = \begin{pmatrix}
         \id & 0 \\ - B & \id
      \end{pmatrix}
      \begin{pmatrix}
         0 & g^{-1} \\ g & 0
      \end{pmatrix}
      \begin{pmatrix}
         \id & 0 \\ B & \id
      \end{pmatrix}~,
\end{equation}
where $g$ is a metric on $X$ and $B$ is a two-form. The patching of $E$
implies $B$ satisfies~\eqref{splitting}, 
so that $B$ may be identified with the curving of the gerbe
used in the twisting of $E$.
Thus the generalized metric $G$ defines a particular splitting of
$E$. In particular, we see from~\eqref{eq:genmetric} that
$G=\me^{-B}G_0\me^B$ where $G_0$ is a generalized metric on $TX\oplus
T^*X$ defined by $g$. 

The generalized metric $G$ naturally encodes the NS fields $g$ and
$B$ as the coset space $O(d,d)/O(d)\times O(d)$. The dilaton $\phi$
appears when one considers the conformal group
$O(d,d)\times\bbR^+$, used to define the generalized spinors
as true polyforms. To define a $O(d)\times O(d)$ substructure in
$O(d,d)\times\bbR^+$, in addition to $G$ which gives the embedding in
the $O(d,d)$ factor, one must give the embedding $\rho$ in the conformal factor
$\rho\in\bbR^+$. Recall that under diffeomorphisms $\rho$
transforms as a section of $\Lambda^dTX$. Given the metric $g$ we can
define the generic embedding by $\rho=\me^{2\phi}/\sqrt{g}$ for
some positive function $\me^{2\phi}$, which we identify as the
dilaton. Note that $\rho$ is by definition invariant under $O(d,d)$
and so one finds the conventional T-duality transformation of the
dilaton under $O(d,d)$. 

Under the generalized Lie derivative, $\Lgen_VG=0$
implies~\cite{Grana:2008yw}
\begin{equation}
      \mathcal{L}_xg = 0~, \qquad
      \mathcal{L}_xB - \dd\lambda = 0 \, ,
\end{equation}
so that $\mathcal{L}_xH=0$ where $H=\dd B$. Such a $V$ is called a
\emph{generalized Killing vector}.

Given $G$ we may decompose generalized spinors in $\Spin(d,d)$ under
$\Spin(d)\times\Spin(d)$. In fact one can go further. Using the projection
$\pi:E\to TX$ the two $\Spin(d)$ groups can be identified and the
generalized spinors may be decomposed as bispinors of $\Spin(d)$:
\begin{equation}
\label{purespinorspinor}
   \Omega_\pm = \me^{-\phi}\me^{-B}\Phi^\text{even/odd} \ . 
\end{equation}
In this expression, one first uses the Clifford map to
identify the bispinors with a generalized spinor 
$\Phi^\text{even/odd}$ of $S_\pm(TX\oplus
T^*X)\cong\Lambda^\text{even/odd}T^*X$ and then uses the splitting $B$
to map to a spinor of $S_\pm(E)$. 
The factor of $\me^{-\phi}$ appears
because the polyforms are really sections of $S^{-1/2}_\pm(E)$
transforming with weight $-\frac{1}{2}$ under conformal rescalings. 
Explicitly, if $\sla{\Phi}$ is a
bispinor, $\Phi\in\Omega^*(X)$ a polyform, and $\gamma^i$ are
$\Spin(d)$ gamma matrices,  the Clifford map is
\bea\label{clifford}
   \sla\Phi = \sum_k\frac{1}{k!}\Phi_{i_1\cdots i_k}\,
      \gamma^{i_1\cdots i_k}
   \quad \longleftrightarrow \quad
   \Phi = \sum_k \frac{1}{k!} \Phi_{i_1\cdots i_k}\,
      \diff x^{i_1}\wedge \cdots \wedge \diff x^{i_k}~.
\eea
The $\Cliff(d,d)$ action is realized via left and right multiplication
by the gamma matrices $\gamma^i$.
For the chiral spinors $\Omega_\pm$ the sum is over $k$ even/odd
respectively. We also note here the Fierz identity
\bea\label{eq:Fierz}
\sla\Phi
   = \frac{1}{n_d}\sum_k \frac{1}{k!} \mathrm{Tr}\left(\sla\Phi
            \gamma_{i_k\cdots i_1} \right)
         \gamma^{i_1\cdots i_k}~,
\eea
where the $\gamma^i$ are $n_d\times n_d$ matrices. Finally, the
generalized metric also defines an action $\Gstar$ on generalized
spinors which is the analogue of the Hodge star. It is given by
\begin{equation}
   \Omega_\pm \mapsto \Gstar\Omega_\pm =
   \me^{-B}\star\lambda(\me^B\Omega_\pm)~,
\end{equation}
where $\lambda$ is the operator defined in~\eqref{eq:lambda} and $\star$
denotes the ordinary Hodge star for the metric $g$.

If $d=2n$ one can also introduce a \emph{generalized almost complex
structure} on $E$. This is a map $\mathcal{J}:E\to E$ with
$\mathcal{J}^2=-\id$ and
$\mukai{\mathcal{J}U}{\mathcal{J}V}=\mukai{U}{V}$ and gives a
decomposition
\be
E_\bbC=L\oplus\bar{L}~, \label{llbdec}
\ee
where $L$ denotes the $+\ii$ eigenspace of $\mathcal{J}$.
Note that $L$ is maximally isotropic:
$\mukai{U}{V}=\mukai{\mathcal{J}U}{\mathcal{J}V}=\mukai{\ii
U}{\ii V}=-\mukai{U}{V}=0$. This defines a $U(n,n)\subset O(2n,2n)$
structure on $E$. By definition
$\mukai{U}{\mathcal{J}V}+\mukai{\mathcal{J}U}{V}=0$, so $\mathcal{J}$ can be
viewed either as an element of $O(2n,2n)$ or of the Lie algebra
$o(2n,2n)$. A generic $\mathcal{J}$ can be written locally as
\begin{equation}
\label{eq:componentJ}
\mathcal{J} =
    \begin{pmatrix}
        I& P\\Q&-I^*
    \end{pmatrix}~,
\end{equation}
where $I^*$ is the linear map on $T^*X$ dual to the map $I$ on $TX$,
$P$ is a bivector and $Q$ is a two-form.
If the twisting~\eqref{eq:S-patch} is trivial, so
$E=TX\oplus T^*X$, there are two canonical examples of generalized
almost complex structures. The first is an ordinary almost complex
structure $I$ on $TX$, for which\footnote{\label{f2} Note that we have
  chosen the opposite sign in \reef{complexcase} compared with
  \cite{gualtieri}. This is so that the $+\ii$ eigenspace is 
  identified with $T^{(1,0)}\oplus T^{*(0,1)}$.} 
\bea\label{complexcase}
\mathcal{J}_1 = \left(\begin{array}{cc} I & 0 \\ 0 &
      -I^*\end{array}\right)~.
\eea
The second is a non-degenerate (stable) two-form $\omega$, for which
\bea\label{symplecticcase}
\mathcal{J}_2 = \left(\begin{array}{cc} 0 & \omega^{-1} \\ -\omega & 0\end{array}\right)~.
\eea
If $\dd\omega=0$ this corresponds to a symplectic structure.

More
generally, a generalized almost complex structure $\mathcal{J}$ is
\emph{integrable} if $L$ is closed under the Courant bracket. That is,
given $U,V\in \SC L$ then $[U,V]\in \SC L$. In the above two cases
(\ref{complexcase}), (\ref{symplecticcase}), this reduces to
integrability of $I$ and the closure of $\omega$,
respectively.
A generalized almost complex structure is equivalent to (the conformal class of) a
\emph{pure spinor} $\Omega$, which simply means a chiral complex
generalized spinor such that the annihilator
\begin{equation}
   L_\Omega = \left\{ U \in {E_\bbC} : U\cdot \Omega = 0 \right\}
\end{equation}
is maximal isotropic. The sub-bundle $L$ defined by $\mathcal{J}$ is
then identified with $L_\Omega$. Notice that $L_{\Omega}$
is invariant under conformal rescalings 
$\Omega\mapsto f \Omega$, 
for any function $f$. A
generalized almost complex structure is therefore more precisely equivalent
to the \emph{pure spinor line bundle} generated by $\Omega$.
Integrability of $\mathcal{J}$ can be expressed as the condition
$\dd\Omega=V\cdot\Omega$ for some $V\in \SC E$. If one can find a nowhere vanishing globally defined $\Omega$
then one has an $SU(n,n)$ structure and if in addition
$\dd\Omega=0$ then one has  a
\emph{generalized Calabi-Yau} structure in the sense\footnote{Note that a different definition is used in \cite{gualtieri}.}  of \cite{hitchin}.
For example, in the complex
structure case (\ref{complexcase}) one has $\Omega =
c\bar {\Omega}_{(n,0)}$, where $\Omega_{(n,0)}$ is the holomorphic
$(n,0)$-form and $c$ is a non-zero constant (the reason why 
$\bar {\Omega}_{(n,0)}$ appears, rather than ${\Omega}_{(n,0)}$,
is directly related to 
the comment in footnote~\ref{f2}).

A generalized vector $V=x+\lambda$ is called (real) \emph{generalized
holomorphic} if $\Lgen_V \mathcal{J}=0$. Equivalently,
$\Lgen_V$ preserves the spinor line bundle generated
by the corresponding pure spinor $\Omega$; that is, $\Lgen_V\Omega=f\Omega$
for some function $f$.

Given a splitting $\Bc$, one can define the corresponding 
generalized complex objects on $TX\oplus T^*X$. In particular, if
$\mathcal{J}$ is the generalized almost complex structure for a pure
spinor $\Omega$, then the corresponding generalized almost complex
structure on $TX\oplus T^*X$ is defined in terms of the annihilator of 
\bea\label{BtransformOmega}
   \Omega^{\Bc} = \me^\Bc \Omega
\eea
and is given by
\bea\label{eq:BtransformJ}
   \mathcal{J}^\Bc \equiv \me^\Bc \mathcal{J}\me^{-\Bc}~.
\eea
In particular, integrability of $\mathcal{J}$ is equivalently to
integrability of $\mathcal{J}^\Bc$ using the twisted Courant
bracket~\eqref{eq:twistedCourant}, or equivalently
$\dd_H\Omega^\Bc=V\cdot\Omega^\Bc$.

Viewing $\mathcal{J}$ as a Lie algebra element one can define its
action on generalized spinors via the Clifford
action~\cite{cavalcanti}. Explicitly, one has
\begin{equation}
    \label{eq:Jaction}
    {\cal J}\cdot = \frac12 \Big(Q_{mn}{\dd x^m\wedge\dd x^n \wedge{}}
    + I^m{}_n[i_{\del_m},\dd x^n\wedge{}]
    + P^{mn}i_{\del_m}i_{\del_n}\Big)~.
\end{equation}
Note that for any generalized vector $V$ one has, under the Clifford
action, $[\mathcal{J}\cdot,V\cdot]=(\mathcal{J}V)\cdot$. 
One can also define the operator $\mathcal{J}_h:S_\pm(E)\to S_\pm(E)$
\begin{equation}\label{Jhitchin}
   \mathcal{J}_h \equiv \me^{\frac{1}{2}\pi\mathcal{J}\cdot}~,
\end{equation}
which is the spinor space representation of $\mathcal{J}$ as an
element of the group $\Spin(d,d)$. If $n$ is even and the pure spinor is a
section of $S_\pm(E)$, then $\mathcal{J}_h$ defines a complex
structure on $S_\mp(E)$, while if $n$ is odd it defines a complex
structure on $S_\pm(E)$. Observe that for any generalized vector $V$ we have the Clifford action identity
$\mathcal{J}_h\cdot V \cdot \mathcal{J}_h^{-1}\cdot =(\mathcal{J}V)\cdot\, $.

Finally, a pair of generalized almost complex structures $\mathcal{J}_1$ and
$\mathcal{J}_2$ are said to be \emph{compatible} if
\bea
[\mathcal{J}_1,\mathcal{J}_2]=0~, \label{gcgcomp}
\eea
and the combination
\begin{equation}
   G = - \mathcal{J}_1\mathcal{J}_2
\end{equation}
is a generalized metric.
If $\Omega_1$ and $\Omega_2$ are the corresponding pure spinors, (\ref{gcgcomp})
is equivalent to ${\mathcal J}_1\cdot \Omega_2={\mathcal
  J}_2\cdot\Omega_1=0$% or $\mukai{\Omega_1}{V\cdot\Omega_2}=0$ for all $V\in\SC{E}$. 
. An example of a pair of compatible
pure spinors is (\ref{complexcase}), (\ref{symplecticcase}), with
the compatibility condition being that $I^{k}_{\ i}\omega_{jk}=g_{ij}$
is positive definite. Note this is $\omega_{ij}=-g_{ik}I^k_{\ j}$,
this mathematics convention differing by a sign to the usual
physics convention. A pair of compatible almost complex structures
defines an $SU(n)\times SU(n)$ structure. A \emph{generalized K\"ahler} structure is an
$SU(n)\times SU(n)$ structure where both generalized almost complex
structures are integrable, while for a \emph{generalized Hermitian} structure
only \emph{one} need be integrable.

Note that an $SU(n)\times SU(n)$ structure can equivalently be specified by
a generalized metric and a pair of chiral $Spin(2n)$ spinors. For example,
for $d=6$ a pair of chiral spinors $\eta^1_+$, $\eta^2_+$
can be used to construct an $SU(3)\times SU(3)$ structure given by
\begin{equation}
\label{purespinorspinor2}
   \Omega_+ = \me^{-\phi}\me^{-B}\eta^1_+ \bar{\eta}^2_+ \, , \qquad
   \Omega_- = \me^{-\phi}\me^{-B}\eta^1_+ \bar{\eta}^2_- \, ,
\end{equation}
with $\eta^2_-\equiv (\eta^2_+)^c$. This will play a central
role in the following sections.
Similarly, for $d=4$ a pair of chiral spinors 
$\eta^1_+$, $\eta^2_+$ give rise to an $SU(2)\times SU(2)$ structure specified 
by two compatible pure spinors, but both of them consist of sums of even forms, 
since now $(\eta^2_+)^c$ is a positive chirality spinor.
We will see such an $SU(2)\times SU(2)$ structure in section 4.
That the spinors have the same
chirality is necessary for them to be compatible in four
dimensions~\cite{gualtierithesis}. 

%%%%%%%%%%%%%%%%%%%%%%%%%%%%%%%%%%%%%%%%%%%%%%%%%%%%%
\section{$AdS_5$ backgrounds as generalized complex geometries}
%%%%%%%%%%%%%%%%%%%%%%%%%%%%%%%%%%%%%%%%%%%%%%%%%%%%%
%%%%%%%%%%%%%%%%%%%%%%%%%%%%%%%%%%%%%%%%%%%%%%%%%%%%%
\subsection{Supersymmetric $AdS_5$ backgrounds}\label{sec:AdS}
%%%%%%%%%%%%%%%%%%%%%%%%%%%%%%%%%%%%%%%%%%%%%%%%%%%%%
Our starting point is the most general class of
supersymmetric $AdS_5$ solutions of type IIB supergravity,
as studied in \cite{Gauntlett:2005ww}. The ten-dimensional metric in Einstein frame is
\bea\label{10dmetric}
g_{E}  =  \me^{2\Delta} \left(g_{{AdS}}  +  g_Y\right)~,
\eea
where $g_Y$ is a Riemannian metric on the compact five-manifold $Y$,
and $\Delta$ is a real function on $Y$. The $AdS_5$ metric $g_{AdS}$ is normalized to have unit radius,
so that
\bea\label{normads}
\mathrm{Ric}_{g_{{AdS}}}  =  -4\, g_{{AdS}}~.
\eea
The ten-dimensional string frame metric is defined to be $g_\sigma\equiv\me^{\phi/2}g_E$.
In addition to the metric, there is the dilaton $\phi$ and NS three-form $H\equiv \dd B$ in the NS sector,
and the forms $F\equiv F_1+F_3+F_5$ in the RR sector. 
The RR fluxes $F_n$ are related to the RR potentials $C_n$ via
\bea\label{RRforms}
F_1 &= & \diff C_0~,\\
F_3 & = & \diff C_2 - H C_0~,\\
F_5 & = & \diff C_4 - H\wedge C_2~.
\eea
These are all taken to be forms
on $Y$, so as to preserve the $SO(4,2)$ symmetry, with the exception of
the self-dual five-form $F_5$ which necessarily takes the form
\bea\label{5flux}
F_5  =  f \left(\vol_{{AdS}} - \widetilde{\vol}_Y\right)~,
\eea
where $f$ is a constant. Here $\widetilde{\vol}_Y$ denotes a volume
form for $(Y,g_Y)$. 
It is related by $\widetilde{\vol}_Y=-\vol_5$ to the
volume form of \cite{Gauntlett:2005ww}, where the latter was given in
terms of an orthonormal frame as $\vol_5=e^{12345}$ and 
used to define, for instance, the Hodge star. In
turns out that, in the Sasaki-Einstein limit, the conventional volume
form is  $\widetilde{\vol}_Y$ rather than $\vol_5$ and so here we will
use the former throughout. In particular, it is the orientation that we
will use when defining integrals over $Y$. 

In \cite{Gauntlett:2005ww} the conditions
for a supersymmetric $AdS_5$ background were written in terms of
two five-dimensional spinors $\xi_1$, $\xi_2$  on $Y$, 
giving the system of equations reproduced here in
(\ref{sone})--(\ref{ssix}) of appendix~\ref{sec:56}. Various spinor
bilinears involving $\xi_1$ and $\xi_2$ were also introduced, and used
to determine the necessary and sufficient conditions for
supersymmetry. For example, it was shown that
\begin{equation}
\label{scalars}
\begin{aligned}
   A &\equiv \tfrac{1}{2}\left(
      \bar{\xi}_1\xi_1 + \bar{\xi}_2\xi_2\right) =1~,\\
   Z &\equiv \bar{\xi}_2\xi_1 =0~.\\
\end{aligned}
\end{equation}
It will be useful for later in this paper to recall the definitions
of the following scalar bilinears:
\begin{equation}
\label{scalarss}
\begin{aligned}
\sin\zeta &\equiv \tfrac{1}{2}\left(
      \bar{\xi}_1\xi_1 - \bar{\xi}_2\xi_2\right)~,\\
   S &\equiv \bar{\xi}^c_2\xi_1~,\\
\end{aligned}
\end{equation}
the one-form bilinears:
\begin{equation}
\label{bilins}
\begin{aligned}
   K &\equiv \bar{\xi}_1^c\beta_{(1)}\xi_2~,\\
   K_3 &\equiv \bar{\xi_2}\beta_{(1)}\xi_1~,\\
   K_4 &\equiv \tfrac{1}{2}\left(
      \bar{\xi}_1\beta_{(1)}\xi_1 - \bar{\xi}_2\beta_{(1)}\xi_2\right) ~,\\
   K_5 &\equiv \tfrac{1}{2}\left(
      \bar{\xi}_1\beta_{(1)}\xi_1 + \bar{\xi}_2\beta_{(1)}\xi_2\right)~,
\end{aligned}
\end{equation}
and the two-form bilinears:
\bea\label{bilins2}
V&=&-\frac{\ii}{2}(\bar\xi_1\beta_{(2)}\xi_1 - \bar\xi_2\beta_{(2)}\xi_2)~,\nn
W&=& -\bar\xi_2\beta_{(2)}\xi_1~.
\eea
Here the $\beta_m$ generate the Clifford algebra for $g_Y$, so
$\{\beta_m,\beta_n\}=2g_{Ymn}$. Equivalently, with respect to any
orthonormal frame, we write $\hat{\beta}_m$ with
$\{\hat{\beta}_m,\hat{\beta}_n\}=2\delta_{mn}$. We have also
introduced the notation $\beta_{(k)}\equiv
\frac{1}{k!}\beta_{m_1\cdots m_k}\diff x^{m_1}\wedge\cdots\wedge\diff
x^{m_k}$. 

A key result of~\cite{Gauntlett:2005ww} is that $K_5^\#$,
the vector dual to the one-form $K_5$, defines a Killing vector that
preserves all of the fluxes. This was identified as corresponding to
the $R$-symmetry in the dual SCFT. Another important result was
\begin{equation}
\label{fxf}
   \me^{-4\Delta}f = 4 \sin\zeta~.
\end{equation}
The Killing spinors $\xi_1$, $\xi_2$ were used to introduce a
canonical five-dimensional orthonormal frame in appendix B of
\cite{Gauntlett:2005ww}, which is convenient for certain
calculations. We will refer to that paper for further
details. Finally, we note that equation (\ref{a}) of appendix
\ref{sec:56} may be used to obtain expressions for the two-form
potentials $B$ and $C_2$ in terms of the bilinear $W$ introduced in
(\ref{bilins2}): 
\bea
B &=& -\frac{4}{f} \me^{6\Delta+\phi/2} \re{W}  + b_2 \;, \label{gaugeb}  \\
C_2 &=& -\frac{4}{f} \me^{6\Delta+\phi/2} C_0 \re{W} - \frac{4}{f}\me^{6\Delta-\phi/2}\im{W}  + c_2 \;.\label{gaugec}
\eea
Here $b_2$ and $c_2$ are real closed two-forms. Notice that the first term
in $B$ in (\ref{gaugeb}) is a globally defined two-form, and thus $H=\diff B$ is
exact. It follows that $[H]=0\in H^3(Y,\R)$, although notice that $b_2$ may
be taken to be globally defined if and only if the \emph{torsion class} of $H$ is
zero in $H^3(Y,\Z)$ (which for simplicity we shall assume). Similar remarks
apply to $C_2$ (up to large gauge transformations of $C_0$).

%%%%%%%%%%%%%%%%%%%%%%%%%%%%%%%%%%%%%%%%%%%%%%%%%%%%%
\subsection{Reformulation as generalized complex geometries}
%%%%%%%%%%%%%%%%%%%%%%%%%%%%%%%%%%%%%%%%%%%%%%%%%%%%%
The supersymmetric $AdS_5$ geometry described above can be simply
reformulated in terms of generalized complex geometry, as in 
the discussion of~\cite{Minasian:2006hv,Butti:2007aq}. The basic
observation is simply that these solutions can be viewed as warped
products of flat four-dimensional space with a six-dimensional
manifold $X$, satisfying a set of supersymmetry conditions that imply
the existence of a particular generalized complex 
geometry~\cite{Grana:2004bg,Grana:2005sn}. As we shall explain in more
detail below, combining this structure with the existence of the
Killing vector $K_5^\#$ precisely generalizes the correspondence 
between Sasaki-Einstein geometry and Calabi-Yau cone geometry. In
the following we analyze this reformulation in detail. We find in
particular that all supersymmetric $AdS_5$ solutions necessarily
satisfy the condition of~\cite{Minasian:2006hv} that there is an
$\SU(2)$-structure on $X$. 

One begins by rewriting the unit $AdS_5$ metric in a Poincar\'e patch as
\bea
g_{{AdS}} =  \frac{\diff r^2}{r^2} +  r^2 g_{\R^{3,1}}~.
\eea
Switching to the string frame, we can consider (\ref{10dmetric})
as a special case of a warped supersymmetric $\R^{3,1}$ solution of the form
\bea\label{stringframe}
g_{\sigma}  =  \me^{2A} g_{\R^{3,1}}  +  g_6~,
\eea
where the warp factor is given by
\bea\label{warp}
\me^{2A}  =  \me^{2\Delta+\phi/2} r^2~,
\eea
and the six-dimensional metric is given by
\be
\label{6metric}
g_6 =  \me^{2\Delta+\phi/2}\left( \frac{\diff r^2}{r^2}  +  g_Y\right)~.
\ee
We also define the six-dimensional volume form as
\begin{equation}
\label{6vol}
   \vol_6\equiv\me^{12\Delta+3\phi}r^5\dd r \wedge \widetilde{\vol}_Y \ . 
\end{equation}
Notice that the six-dimensional manifold $X$, on which $g_6$ is a metric,
is a product $\R^+\times Y$, where $r$ may be interpreted as a coordinate
on $\R^+$. In particular, $X$ is non-compact.
It thus follows that supersymmetric $AdS_5$ solutions are
special cases of supersymmetric $\R^{3,1}$ solutions.

In \cite{Grana:2005sn} the general conditions
for an $\mathcal{N} =1$ supersymmetric $\R^{3,1}$ background, in the
string frame metric (\ref{stringframe}), were written in terms
of two chiral six-dimensional spinors $\eta^1_+$, $\eta^2_+$ on $X$,
namely the system of equations given here in (\ref{eq:intgravB})--(\ref{swap}).
The relation between the two sets of Killing spinors, for $AdS_5$ solutions,  is given by
first decomposing $\mathrm{Cliff}(6)$ into $\mathrm{Cliff}(5)$
via
\bea\label{6to5}
\hat{\gamma}_{m}&=&\hat{\beta}_{m}\otimes \sigma_3~,\qquad m=1,\ldots,5\nn
\hat{\gamma}_6&=&1\otimes\sigma_1~,
\eea
where $\hat{\gamma}_i$, $i=1,\ldots,6$, generate $\mathrm{Cliff}(6)$
and
$\sigma_\alpha$, $\alpha=1,2,3$, denote the Pauli matrices. Changing basis
to $\xi_1=\chi_1+\ii\chi_2$, $\xi_2=\chi_1-\ii\chi_2$, we then have
\bea\label{Aaron}
\eta_+^1 & = & \me^{A/2} \left(\begin{array}{c} \chi_1 \\ \ii\chi_1 \end{array}\right),
\qquad
\eta_-^1  =  \me^{A/2} \left(\begin{array}{c} -\chi_1^{c} \\ \ii\chi_1^{c} \end{array}\right)
\nonumber\\
\eta_+^2 & = & \me^{A/2} \left(\begin{array}{c} -\chi_2 \\ -\ii\chi_2\end{array}\right),
\qquad
\eta_-^2  =  \me^{A/2} \left(\begin{array}{c} \chi_2^{c} \\ -\ii\chi_2^{c} \end{array}\right)~,
\eea
where $\chi_i^c \equiv \tilde{D}_5 \chi_i^*$ denotes 5D charge conjugation,
and correspondingly $\eta_-^i\equiv (\eta_+^i)^c\equiv D_6(\eta_+^i)^*$ where
$D_6=\tilde{D}_5\otimes \sigma_2$. For
further details, see appendix \ref{sec:56}.
Using the two chiral spinors $\eta_1^+$, $\eta_2^+$ we may  define the bispinors
\bea\label{Phis}
\Phi_+ \, \equiv \, \eta_+^{1} \otimes \bar{\eta}_+^{2}, \qquad \Phi_- \, \equiv \, \eta_+^1 \otimes
\bar{\eta}_-^{2}~.
\eea
Notice that, in the conventions of appendix \ref{sec:56}, we have $A_6=1$, so that
$\bar{\eta}\equiv \eta^\dagger$
is just the Hermitian conjugate.
Via the Clifford map (\ref{clifford}) the bispinors for
$\Spin(6)$ in (\ref{Phis}) may also be viewed as elements of
$\Omega^*(X,\C)$. We will mainly tend to think of $\Phi_\pm$ as complex
differential forms of mixed degree. These are then $\Spin(6,6)$ spinors,
as explained in section \ref{sec:genmetric}.
In fact $\Phi_\pm$ in (\ref{Phis}) are both \emph{pure} spinors,
and also \emph{compatible}. They then
define an $SU(3)\times SU(3)$ structure on $TX\oplus T^*X$.

In terms of (\ref{Phis}), the Killing spinor equations
for a general supersymmetric $\bbR^{3,1}$ solution ({\it i.e.} not necessarily associated with
an $AdS_5$ solution, but with vanishing four-dimensional cosmological constant)
may be rewritten as \cite{Grana:2006kf} (see also \cite{Grana:2004bg})
\bea\label{integrable}
\diff_H \left(\me^{2A-\phi}\, \Phi_-\right) & = & 0~,\\\label{notintegrable}
\diff_H \left(\me^{2A-\phi}\, \Phi_+\right) &=  & \me^{2A-\phi}\diff
A\wedge \bar{\Phi}_+ \nn
&&+\, \frac{1}{16}\me^{2A}\left[(|a|^2-|b|^2)F  +  \ii (|a|^2 + |b|^2)
   \star \lambda(F)\right]~.   \label{mink2}
\eea
Here recall that $F= F_1 + F_3  +  F_5$ is the sum of RR fields and from (\ref{eq:lambda})
\bea\label{lambdaF}
\lambda(F)  = F_1  -  F_3  +  F_5~.
\eea
Note that the Hodge star is with respect to the metric $g_6$, with positive orientation given by
$\dd r \wedge \widetilde{\vol}_Y$.
The remaining Bianchi identities and equations of motion
are ({\it cf}.~\cite{Grana:2006kf} equations (4.9)--(4.10))
\bea \label{Bianchi}
&&\dd H = 0\; ,\qquad\diff_H F  =  \delta_{\text{source}}\;, \\
&&\dd(\me^{4A-2\phi} \star H) -\me^{4A} F_n \wedge \star F_{n+2}=0~,\\
&&(\dd+H\wedge{}) (\me^{4A} \star F) = 0~.
\eea
The equation of motion for $F$ can also be written as
\bea
\dd \left[ \me^{4A} \me^{-B}\star \lambda \left(F\right) \right] = 0~, \label{susyom}
\eea
and follows from the supersymmetry equations.
In fact, for $AdS_5$ solutions it was shown in \cite{Gauntlett:2005ww}
that supersymmetry implies \emph{all} of the equations of motion and Bianchi
identities. We have also introduced the spinor norms
\bea
|a|^2  =  |\eta_1^+|^2~, \qquad |b|^2 =  |\eta_2^+|^2~,
\eea
which for a supersymmetric $\bbR^{3,1}$ background must satisfy
\be \label{normsab}
|a|^2  +  |b|^2  =    \me^A c_+~,  \qquad |a|^2  -  |b|^2  =   \me^{-A} c_-  \;,
\ee
where $c_{\pm}$ are constants. Upon squaring and subtracting the equations one obtains
\be\label{norms}
\left|\Phi_{\pm}\right|^2  =  \frac{1}{8}\, |a|^2|b|^2 = \frac{1}{32}\left(\me^{2A} c^2_+  -   \me^{-2A} c_-^2\right)  \;.
\ee

As we now show, for the particular case of $AdS_5$ solutions the above
equations simplify somewhat.
In this case it is possible to fix the constant $c_-$ in (\ref{norms})
by the scaling of $\Phi_{\pm}$ with $r$ which, using
(\ref{warp}), implies that $c_-=0$ and hence $|\eta_1^+|=|\eta_2^+|$.
This is consistent with the equation $Z=0$ in (\ref{scalars}), since from (\ref{Aaron}) we see that $|\eta^1_\pm|=|\eta^2_\pm|$
is equivalent to $\mathrm{Re}\, Z =0$. Notice that $c_-=0$
is also a necessary condition in order to have supersymmetric probe
branes \cite{Martucci:2005ht}. 
The normalization that was used in \cite{Gauntlett:2005ww} implies
$|a|^2=|b|^2=\me^A$ and hence $c_+=2$.
One can actually go a little further. In~\cite{Minasian:2006hv} it was
assumed that there was an $\SU(2)$-structure on the cone. In terms of
the spinors $\eta_i^+$ this is equivalent to the condition that, in
addition to $c_-=0$, one has
\begin{equation}
   \bar{\eta}_1^+\eta_2^+ + \bar{\eta}_2^+\eta_1^+ = 0 \ . 
\end{equation}
However it is easy to see that this is equivalent to $\im Z=0$, which
again is required by supersymmetry on $Y$. Thus in fact all supersymmetric
$AdS_5$ solutions necessarily satisfy the $\SU(2)$ condition
of~\cite{Minasian:2006hv}. 

We now define the pure spinor
\bea\label{ps1def}
\Omega_-^B  \equiv  \me^{2A-\phi}  \Phi_-~,
\eea
which by \eqref{integrable} is $\dd_H$ closed, $\diff_H \Omega_-^B  =  0$.
The associated generalized almost complex structure $\mathcal{J}^B_-$ is
then integrable with respect to the twisted Courant bracket
 (\ref{eq:twistedCourant}).
We also define
\bea\label{relomphi}
\Omega_-  \equiv  \me^{-B} \Omega_-^B  =  \me^{-B} \me^{2A-\phi} \Phi_-~,
\eea
which is closed under the usual exterior derivative:
\bea
\diff \Omega_-  =  0~.
\eea
The associated generalized almost complex structure, which we denote by
$\mathcal{J}_-$,\footnote{The generalized
complex structures $\mathcal{J}_-^B$ and $\mathcal{J}_-$ are related by (\ref{eq:BtransformJ}).} 
is then integrable. 
Combined with the fact that the norm of $\Phi_-$, and hence of $\Omega_-$, is nowhere vanishing,
this means, in particular, that we have a generalized Calabi-Yau manifold in the sense of \cite{hitchin}.

We similarly define
\bea\label{Omegaplus}
\Omega_+ \equiv   \me^{-B} \me^{2A-\phi} \Phi_+~.
\eea
However, the corresponding
generalized almost complex structure
$\mathcal{J}_+$ is not integrable in general, its integrability being obstructed
by the RR fields in \eqref{notintegrable}. If it \emph{were} integrable,
we would have a generalized K\"ahler manifold.
With these definitions we can write the supersymmetry equations for $AdS_5$ solutions
as
\bea
\dd \Omega_- &=& 0 \;, \nn
\dd \Omega_+ &=& \dd A \wedge \bar{\Omega}_+ + \frac{\ii}{8}\me^{3A}\me^{-B} \star
    \lambda \left(F\right)~.
\eea
It is worth noting that the latter equation may also be written as
\bea
\label{ReSUSY}
\dd \left(\me^{-A}\mathrm{Re}\, \Omega_+\right) & = & 0~,\\\label{ImSUSY}
\dd \left(\me^A\mathrm{Im}\, \Omega_+\right)&=& \frac{1}{8}\me^{4A}\me^{-B}\star\lambda \left(F\right)~,
\eea
and that in turn equation (\ref{ImSUSY}) can be written as \cite{Tomasiello:2007zq}
\be
\me^{-B}F = 8 {\cal J}_-\cdot \dd\left( \me^{-3A} \im{\Omega}_+ \right) = 8 \dd^{\cal J_-}  \left( \me^{-3A} \im{\Omega}_+ \right) \;, \label{djomp}
\ee
where $\dd^{\cal J_-}\equiv -[\dd,{\cal J}_-\cdot]$.
For most of the paper we will demand that $F_5\neq0$, or equivalently $f\ne 0$. Physically
this corresponds to having non-vanishing D3-brane charge. It would be interesting to know
whether or not all supersymmetric $AdS_5$ solutions of type IIB supergravity have this property.
%%%%%%%%%%%%%%%%%%%%%%%%%%%%%%%%%%%%%%%%%%%%%%%%%%%%%
\subsection{Canonical vector fields}
%%%%%%%%%%%%%%%%%%%%%%%%%%%%%%%%%%%%%%%%%%%%%%%%%%%%%
In this section we examine the geometric properties of the generalized vector
fields $r\partial_r$,
$\xi\equiv \mathcal{J}_-(r\de_r)$ and $\eta \equiv  \mathcal{J}_- (\dd \log r)$.
As in the Sasaki-Einstein case, $r\partial_r$ and $\xi$ correspond respectively to the
dilatation symmetry and the $R$-symmetry in the dual SCFT (while $\eta$ is related to a contact structure on $Y$, 
as we shall show later in section \ref{ompgenplus}).
%%%%%%%%%%%%%%%%%%%%%%%%%%%%%%%%%%%%%%%%%%%%%%%%%%%%
\subsubsection{Dilatation symmetry}
%%%%%%%%%%%%%%%%%%%%%%%%%%%%%%%%%%%%%%%%%%%%%%%%%%%%%
We begin with the dilatation vector field $r\de_r$.
It immediately follows from (\ref{warp}), (\ref{Aaron})  and
(\ref{Phis}) that
\be\label{Phidimension1}
\mathcal{L}_{r \de_r} \Phi_{\pm}  =  \Phi_{\pm} \;,
\ee
and therefore
\be\label{Omegadimension3}
\mathcal{L}_{r \de_r} \Omega_\pm  =  3 \Omega_\pm~.\ee
This follows since $\me^{2A}$ has scaling dimension 2 (\ref{warp}), and both
the $B$-field and the dilaton $\phi$ are pull-backs from $Y$.
Notice that equation (\ref{Omegadimension3}) may also be trivially rewritten in terms of the generalized Lie derivative (\ref{eq:genLS}):
\begin{equation}\label{Omega3}
   \Lgen_{r\del_r}\Omega_\pm
       = 3 \Omega_\pm~.
\end{equation}
This  implies that
\begin{equation}\label{KillBill}
   \Lgen_{r\del_r}\mathcal{J}_\pm = 0~.
\end{equation}
To see this, recall that $\mathcal{J}_\pm$ is
defined by saying that its $+\ii$ eigenspace is equal to
the annihilator $L_{\Omega_\pm}$ of $\Omega_\pm$, and
the latter is clearly preserved under the one-parameter
family of (generalized) diffeomorphisms generated by $r\de_r$. It
further follows that $\Lgen_{r\del_r}G=0$, where $G$ is the
generalized metric $G=-{\cal J}_+{\cal J}_-= 
-{\cal J}_+{\cal J}_-$,
so that $r\del_r$ is
generalized Killing. Equation (\ref{KillBill})
says  that $r \de_r$ is a (real)
\emph{generalized holomorphic} vector field for
the integrable generalized complex structure $\mathcal{J}_-$.
We shall not use this terminology for $\mathcal{J}_+$,
since the latter is not in general integrable. Clearly, this generalizes the Sasaki-Einstein
result where the cone is Calabi-Yau and the dilatation
vector $r\del_r$ is holomorphic.

%%%%%%%%%%%%%%%%%%%%%%%%%%%%%%%%%%%%%%%%%%%%%%%%%%%%%
\subsubsection{$R$-symmetry}
\label{sec:rsymmetry}
%%%%%%%%%%%%%%%%%%%%%%%%%%%%%%%%%%%%%%%%%%%%%%%%%%%%%
We next define the generalized vectors
\bea
\xi &\equiv&  \mathcal{J}_- (r\de_r)  \;, \\
\eta &\equiv&  \mathcal{J}_- (\dd \log r)  \;, \label{etadef}
\eea
which are, in general, a mixture of vectors and one-forms.
Recall that the generalized almost complex
structures $\mathcal{J}_\pm$ are related to the generalized metric via
$G=-\mathcal{J}_+\mathcal{J}_-=-\mathcal{J}_-\mathcal{J}_+$. The
conical form~\eqref{6metric} of the metric $g_6$ and the fact that $B$
has no component along $\dd r$ implies that
$   G \; \dd \log r = \me^{-2\Delta-\tfrac{\phi}{2}}r\del_r$,
$G \; r\del_r = \me^{2\Delta+\tfrac{\phi}{2}} \dd\log r$,
and hence in addition to (\ref{etadef}) we may also write
\bea
\xi &=& \me^{2\Delta+\tfrac{\phi}{2}} \mathcal{J}_+ (\dd\log r)~,\nn
\eta &=&  \me^{-2\Delta-\tfrac{\phi}{2}} \mathcal{J}_+ (r\de r)~.
\eea
We may split $\xi$ and $\eta$ into a vector part and a one-form part, in a fixed splitting of $E$,
\bea
\xi  &=& \xi_v + \xi_f \;, \\
\eta &=& \eta_v + \eta_f \;.
\eea
By carrying out a calculation, presented in appendix \ref{sec:Rsymappendix}, we may then write these as bilinears constructed from the five-dimensional
Killing spinors (\ref{bilins}):
\bea\label{tranresvects}
\xi_v& =& K_5^{\#}~,\nn
\xi_f&=&i_{\xi_v}b_2~,\nn
\eta_v &=& \me^{-2\Delta - \phi/2}\re{K_3^{\#}}~,\nn
\eta_f&=&\frac{4}{f} \me^{4\Delta} K_4 + i_{\eta_v} b_2~.
\eea
As discussed in appendix \ref{sec:Rsymappendix}, it is the $B$-transform, $\xi^B$, of the generalized vector $\xi$ that is naturally related to
the bilinears of \cite{Gauntlett:2005ww}. We have obtained \reef{tranresvects} by performing an inverse $B$-transform
using the expression for the $B$-field given in terms of bilinears presented in \reef{gaugeb}. In particular, this is where
the closed two-form $b_2$ appears.
Since the $B$-transform of $b_2$ by an exact form is a generalized diffeomorphism, and a gauge symmetry of
string theory, we see that the physical information in $b_2$ is represented by its cohomology class in $H^2(X,\R)$.
More precisely, large gauge transformations of the $B$-field, which correspond to tensoring the underlying
gerbe by a unitary line bundle on $X$, lead to the torus $H^2(X,\R)/H^2(X,\Z)$ (with suitable normalization).
Turning on the two-form $b_2$ corresponds to giving vacuum expectation values to moduli (of the NS field $B$) and so is a symmetry of the supersymmetry equations. It is therefore left undetermined. In the field theory dual, the cohomology
class of $b_2$ thus corresponds to a marginal deformation.

In \cite{Gauntlett:2005ww} it was shown that $K_5^\#$ is a Killing vector that
preserved all of the fluxes, and thus $K_5^\#$ was identified as being dual to the $R$-symmetry in the SCFT.
In the generalized geometry we can show the stronger conditions that 
\be
\Lgen_\xi\mathcal{J}_\pm=0~,
\ee
and hence $\xi$ is a generalized
holomorphic Killing vector field. In fact it is straightforward to show $\Lgen_{\xi}\FF =-3\ii \FF$
and hence $\Lgen_\xi{\mathcal J}_-=0$. Indeed since $\dd\Omega_-=0$ and $r\del_r - \ii\xi\in L_{\Omega_-}$ annihilates
$\FF$,  using \reef{eq:genLS} and (\ref{Omega3}) we have
\begin{equation}\label{charged}
   \Lgen_{\xi}\FF = \dd\left( \xi\cdot\FF\right)
       = -\ii \dd\left( r\de_r \cdot\FF\right)
       = -\ii \Lgen_{r\del_r}\FF
       = -3\ii \FF~.
\end{equation}
In appendix \ref{sec:Rsymappendix} we show that $\Lgen_{\xi}\Omega_+ =0$ and hence $\Lgen_\xi{\mathcal J}_+=0$.
There we also show that
\be\label{killthosefluxes}
\Lgen_\xi(\me^{-B}F)=0~.
\ee
Thus, we have established that $\xi\equiv\mathcal{J}_-(r\del_r)$ is a generalized holomorphic vector field, which
moreover is generalized Killing for the generalized metric $G=-\mathcal{J}_-\mathcal{J}_+$, and also
preserves the RR fluxes.
Again, this clearly generalizes the Sasaki-Einstein result, where $\xi=I(r\del_r)$ is
a holomorphic Killing vector field for the Calabi-Yau cone.

To conclude this section we note that when $f\ne 0$ the vector field $\xi_v=K_5^\#$ is nowhere vanishing on $Y=\{r=1\}$. 
One can see this from the formula
\bea
|K_5^\#|^2 = \sin^2\zeta+|S|^2~,
\eea
and using \reef{fxf}. Thus for $f\neq 0$, $\xi_v$ acts locally freely 
on $Y$ and hence the orbits of $\xi_v$ define a corresponding one-dimensional foliation of 
$Y$. This is again precisely as in the Sasaki-Einstein case (although in the Sasaki-Einstein case the norm of $\xi_v$ is 
constant).

%%%%%%%%%%%%%%%%%%%%%%%%%%%%%%%%%%%%%%%%%%%%%%%%%%%%%%%%%%%%%%%%%%%%%%%%%%
 
\section{Generalized reduction of $AdS_5$ backgrounds}
\label{phipgen}
%%%%%%%%%%%%%%%%%%%%%%%%%%%%%%%%%%%%%%%%%%%%%%%%%%%%%%%%%%%%%%%%%%%%%%%%%%
Recall that in the Sasaki-Einstein case
one can consider the symplectic reduction of the Calabi-Yau cone metric with respect to the 
$R$-symmetry Killing vector $\xi$ (or equivalently a holomorphic
quotient with respect to $r\del_r-\ii\xi$). Generically $\xi$ does
not define a $U(1)$ fibration and the four-dimensional reduced space is not a
manifold. Nonetheless, locally one can consider the geometry on the
transversal section to the foliation formed by the orbits of $\xi$ in
the Sasaki-Einstein space. The result of the reduction is that this 
four-dimensional geometry is K\"ahler-Einstein. Thus locally one can
always write the Sasaki-Einstein metric as  
\begin{equation}
\label{eq:KEred}
   g_Y = \eta\otimes\eta + g_\text{KE}
\end{equation}
where $g_\text{KE}$ is a K\"ahler-Einstein metric. 

The existence of the generalized holomorphic vectors $\xi$ and
$r\del_r$ in the generic case suggests one can make an analogous
generalized reduction to four dimensions. In this section, we show
that this is indeed the case following the theory of generalized
quotients developed in~\cite{bcg1,bcg}. We first review the formalism and
then apply it to our particular case, showing that there is a
generalized Hermitian structure on the local transversal section,
giving the conditions satisfied by the corresponding reduced pure
spinors.  

%%%%%%%%%%%%%%%%%%%%%%%%%%%%%%%%%%%%%%%%%%%%%%%%%%%%%%%%%%%%%%%%%%%%%

\subsection{Generalized reductions}

We will follow the description of generalized quotients given
in~\cite{bcg}\footnote{Note that the bracket $[\![,]\!]$ used in~\cite{bcg}
  is the Dorfman bracket or generalized Lie derivative
  $[\![V,W]\!]=\Lgen_VW$ and is not anti-symmetric.}. These include
both symplectic reductions and complex quotients as special cases. One
first needs to introduce the \emph{reduction data}. 

In conventional geometry, the action of a Lie group $G$ on $M$ is
generated infinitesimally by a set of vector fields, defined by a map
from the Lie algebra $\psi:\mathfrak{g}\to\SC{TM}$. Given a vector
field $x\in\SC{TM}$, the infinitesimal action of $u\in\mathfrak{g}$ is 
then just the Lie derivative (or in this case Lie bracket) 
\begin{equation}
   \delta x = \mathcal{L}_{\psi(u)}x = [ \psi(u), x] \ . 
\end{equation}
One requires that given $u,v\in\mathfrak{g}$, one has 
$[\psi(u),\psi(v)]=\psi([u,v])$ so that
\begin{equation}
\label{Liehomo}
   \mathcal{L}_{\psi(u)}\mathcal{L}_{\psi(v)}
      - \mathcal{L}_{\psi(v)}\mathcal{L}_{\psi(u)}
      = \mathcal{L}_{[\psi(u),\psi(v)]}
      = \mathcal{L}_{\psi([u,v])} \ , 
\end{equation}
and thus there is a Lie algebra homomorphism between $\mathfrak{g}$ and
the algebra of vector fields under the Lie bracket. 

In generalized geometry, we have a larger group of symmetries,
diffeomorphisms and $B$-shifts, which are generated infinitesimally by
the generalized Lie derivative~\eqref{eq:Lgen}. Thus given an action of
$G$ on $M$, it is natural to consider the infinitesimal ``lifted
action'' of $G$ on $E$ defined by the map
$\tpsi:\mathfrak{g}\to\SC{E}$, such that for any $V\in\SC{E}$ and 
$u\in\mathfrak{g}$ we have 
\begin{equation}
   \delta V = \Lgen_{\tpsi(u)} V \ , 
\end{equation}
and under the projection $\pi:E\to TM$ we simply get the vector fields
$\psi(u)$, that is 
\begin{equation}
   \pi\tpsi(u) = \psi(u) \ . 
\end{equation}
Such transformations are infinitesimal automorphisms of $E$ that have
the property that they preserve both the metric $\mukai{\cdot}{\cdot}$
on $E$ and the Courant bracket~\eqref{eq:Courant}. If we again assume
that  
\begin{equation}
   \Lgen_{\tpsi(u)}\Lgen_{\tpsi(v)}
      - \Lgen_{\tpsi(v)}\Lgen_{\tpsi(u)}
      = \Lgen_{\tpsi([u,v])} \ , 
\end{equation}
then $\tpsi$ defines an equivariant structure on $E$. (Note that this
is equivalent to the Courant bracket condition
$[\tpsi(u),\tpsi(v)]=\tpsi([u,v])$.) In what follows it will also be
assumed that $\tpsi$ is isotropic, that is 
\begin{equation}
   \mukai{\tpsi(u_1)}{\tpsi(u_2)}=0 
\end{equation}
for all $u_1,u_2\in\mathfrak{g}$. 

One can actually define a more general action on $E$ which is a
homomorphism between algebras with Courant brackets rather than Lie
algebras. One starts by extending $\mathfrak{g}$ to a larger
algebra. The construction considered in~\cite{bcg} which is relevant for us is as follows. Let
$\mathfrak{h}$ be a vector space on which there is some 
representation of $\mathfrak{g}$. Then we can form a ``Courant
algebra'' $\mathfrak{a}=\mathfrak{g}\oplus\mathfrak{h}$ with Courant
bracket\footnote{Note we take a slightly different definition of the
  bracket to that in~\cite{bcg} in to order to match the Courant 
  bracket~\eqref{eq:Courant} on $E$.} 
given $u_i\in\mathfrak{g}$ and $w_i\in\mathfrak{h}$
\begin{equation}
\label{eq:hemisemi}
   [ (u_1,w_1),(u_2,w_2) ] 
      = (\, [u_1,u_2] \, , \,
          \tfrac{1}{2}(u_1\cdot w_2 - u_2\cdot w_1) \, )~,
\end{equation}
where $u\cdot w$ is the action of $\mathfrak{g}$ on
$\mathfrak{h}$. Suppose in addition $\mu:M\to\mathfrak{h}^*$ is a
$\mathfrak{g}$-equivariant map, meaning
$\mathcal{L}_{\psi(u)}\mu(w)=\mu(u\cdot w)$ for all $u\in\mathfrak{g}$
and $w\in\mathfrak{h}$. Then, given some isotropic lifted action
$\tpsi$ of $\mathfrak{g}$, one can then define the \emph{extended
  action} 
\begin{equation}
\label{Psi-map}
\begin{aligned}
   \Psi : \mathfrak{g}\oplus\mathfrak{h} &\to \SC{E} \ , \\
   (u,w) &\mapsto \tpsi(u) + \dd \mu(w) \ , 
\end{aligned} 
\end{equation}
which, it is easy to show, has the property that 
\begin{equation}
   [ \Psi(u_1,w_1), \Psi(u_2,w_2) ] 
      = \Psi([u_1,u_2],\tfrac{1}{2}(u_1\cdot w_2 - u_1\cdot w_2)) \ . 
\end{equation}
and hence $\Psi$ defines a
homomorphism of Courant algebras as opposed to Lie algebras as
in~\eqref{Liehomo}. Note that the extra factor $\dd\mu(w)$
in~\eqref{Psi-map} corresponds to a trivial $B$-shift, and thus
$\Lgen_{\Psi(u,v)}=\Lgen_{\tpsi(u)}$. Note also that $\mu$ will
play the role of a moment map. In the conventional case of symplectic
reductions one has  $\mu:M\to\mathfrak{g}^*$, whereas here
$\mathfrak{h}$ can be any representation space. The triple
$(\tpsi,\mathfrak{h},\mu)$ is known as the \emph{reduction data}. 

This reduction data can then be used to define a reduced generalized
tangent bundle $\Ered$. First one makes the usual assumptions about
$\mu$ and the $G$ action on $M$ so that $\Mred=\mu^{-1}(0)/G$ is a
manifold. (This requires that $0$ is a regular point of $\mu$ and that
the $G$ action on $\mu^{-1}(0)$ is free and proper.) Then define the
sub-bundle $K$ which is the image of the bundle map
$\mathfrak{a}\times M$ associated to $\Psi$, that is 
\begin{equation}
   K = \left\{ \tpsi(u)+\dd\mu(w), u\in\mathfrak{g}, w\in\mathfrak{h}
         \right\} \subseteq E \ , 
\end{equation}
and also the orthogonal bundle $K^\perp$, the fibres of which are
orthogonal to $K$ with respect to the $O(d,d)$ metric
$\mukai{\cdot}{\cdot}$. One can then construct the generalized tangent
space on $\Mred$ 
\begin{equation}
   \Ered = \left.
      \frac{K^\perp|_{\mu^{-1}(0)}}{K|_{\mu^{-1}(0)}}\right/ G \ . 
\end{equation}
The main results of~\cite{bcg1,bcg} are then to show how various
geometrical structures can be transported from $E$ to $\Ered$. The
case of particular interest to us is that of generalized Hermitian
reduction. 

As discussed in section~\ref{sec:genmetric} a generalized Hermitian
manifold is a generalized complex manifold with a compatible
generalized metric (or equivalently a second, compatible, 
generalized almost complex structure). Let $\mathcal{J}$ be the
integrable generalized complex structure and $G$ be the generalized
metric. Given some reduction data $(\tpsi,\mathfrak{h},\mu)$, 
recalling $\Lgen_{\Psi(u,v)}=\Lgen_{\tpsi(u)}$,
the structures are $G$-invariant if 
\begin{equation}
   \Lgen_{\tpsi(u)} G = \Lgen_{\tpsi(u)} \mathcal{J} = 0 \ , 
\end{equation}
for all $u\in\mathfrak{g}$. One can also define the sub-bundles 
\begin{equation}
   K^G = GK^\perp \cap K^\perp  \ , 
\end{equation}
which is the sub-bundle of $K^\perp$ the fibres of which are
orthogonal to $K$, with respect to $G$, and 
\be
E_K=K\oplus GK
\ee
which is the $G$-orthogonal complement to $K^G$.
Theorem~4.4 of~\cite{bcg} then states
\newtheorem{theorem}{Theorem}
\begin{theorem}[Generalized Hermitian reduction~\cite{bcg}] 
   Let $E$ be a generalized tangent space over $M$ with reduction data
   $(\tpsi,\mathfrak{h},\mu)$. Suppose $E$ is equipped with a
   $G$-invariant generalized Hermitian structure $(\mathcal{J},G)$. If over 
$\mu^{-1}(0)$, $\mathcal{J}K^G = K^G$, or equivalently $\mathcal{J}E_K = E_K$,
   then $\mathcal{J}$ and $G$ can be reduced to $\Ered$ where they
   define a generalized Hermitian structure. 
\end{theorem}
Even if the group action is such that the reduced space is not a manifold, 
one can still define a generalized Hermitian structure on the transversal section to the foliation. 

%%%%%%%%%%%%%%%%%%%%%%%%%%%%%%%%%%%%%%%%%%%%%%%%%%%%%%%%%%%%%%%%%%%%%%%

\subsection{Generalized reduction from $\xi$}
\label{sec:xi-red}

We now use the reduction formalism to show that the generalized
Calabi-Yau geometry on the cone $X$ reduces to a generalized Hermitian
geometry in four dimensions. This is the analogue of the reduced
K\"ahler-Einstein geometry in the Sasaki-Einstein case.

First we note that there is a group action on the cone $X$ generated by
the vectors $r\del_r$ and $\xi_v$. These commute and the corresponding
Lie algebra is simply $\bbR\oplus\bbR$. If the orbits of $\xi_v$ form
a $U(1)$ action then together $r\del_r $ and $\xi_v$ integrate
to a $\bbC^*$ action, but this need not be the case. The generalized
vectors $r\del_r$ and $\xi$ give a lifted action of $\bbR\oplus\bbR$
on $E$, so that, if $u=(a,b)\in\bbR\oplus\bbR$, 
\begin{equation}
   \tpsi(u) = a r\del_r + b \xi \ . 
\end{equation}
By definition we have $\pi\tpsi(u)=ar\del_r+b\xi_v$. Under the
Courant bracket, given the expressions~\eqref{tranresvects} we see that
$[r\del_r,\xi]=0$, and hence 
\begin{equation}
   [ \tpsi(u_1), \tpsi(u_2) ] = 0 = \tpsi([u_1,u_2]) 
\end{equation}
for all $u_1$ and $u_2$, as required for a lifted action. Furthermore,
from~\eqref{isoconst} we see that $\tpsi$ is isotropic. 

We also have a generalized Hermitian structure on $X$ given by $\mathcal{J}_\pm$. The generalized
complex structure $\mathcal{J}_-$ is integrable, and we have the
compatible generalized metric $G=-\mathcal{J}_+\mathcal{J}_-$. In section 3 we showed
$\Lgen_{r\del_r}\mathcal{J}_\pm=\Lgen_\xi\mathcal{J}_\pm=0$ and hence
the Hermitian structure is invariant under both group actions. 

There are then two different ways we can view the generalized
reduction, mirroring the symplectic reduction and the complex quotient
in the Sasaki-Einstein case. In the first reduction, 
we take $\mathfrak{g}=\bbR$ generated by $\xi_v$, and in the reduction
data we take $\mathfrak{h}=\mathfrak{g}$ and $\mu=\log r$. This is the
same moment map one takes in the symplectic reduction. In the second
case we take the complex Lie algebra $\mathfrak{g}=\bbC$ generated by
$r\del_r-\ii\xi_v$ and $\mathfrak{h}=0$ so there is no moment
map. The reduction is then analogous to a complex quotient. We now
discuss these in turn. As in the Sasaki-Einstein case, both lead to
the same reduced structure. 

\subsubsection{$\mathfrak{g}=\bbR$ reduction}
In this case, the reduction data is 
\begin{equation}
   \tpsi(u) = u \xi \ , 
   \qquad 
   \mathfrak{h} = \bbR \ , 
   \qquad 
   \mu = \log r \ . 
\end{equation}
We have already seen that $\tpsi$ is an isotropic lifted action. It is
also clear that $\mu$ is $\mathfrak{g}$-equivariant since,
from~\eqref{isoconst}, $i_{\xi_v}\dd\mu=0$. Thus
$(\tpsi,\mathfrak{h},\mu)$ are suitable reduction data.
Furthermore, $\mathcal{J}_-$ and $G$ are both invariant under $\tpsi(u)$. 
We have 
\begin{equation}
   \mu^{-1}(0) = Y~,
\end{equation}
and given $u\in\mathfrak{g}$ and $v\in\mathfrak{h}$
\begin{equation}
   K = \{ u\xi + v\, \dd \log r \} \ . 
\end{equation}
Using 
\begin{equation}
\label{G-action}
   G \xi = \me^{2\Delta+\phi/2}\eta \ , \qquad 
   G \,\dd\log r = \me^{-2\Delta-\phi/2}r\del_r
\end{equation}
we have 
\begin{equation}
   GK = \{ u'\eta + v'r\del_r \} \ , 
\end{equation}
and hence
\begin{equation}
   E_K \equiv K \oplus GK 
      = \{ u\xi + v\, \dd \log r + u'\eta + v'r\del_r \} \ .
\end{equation}
Using the definitions~\eqref{etadef} we immediately see that
$\mathcal{J}_-E_K=E_K$. Hence, assuming the action of $\xi_v$ on $Y$
gives a $U(1)$ fibration, using the generalized Hermitian reduction
theorem,  we see that we have a generalized Hermitian 
structure on $\Ered$ over the four-dimensional space $\Mred =
Y/U(1)$. More generally, we get a generalized Hermitian structure on the
transversal section to the $\xi_v$ orbits. 

\subsubsection{$\mathfrak{g}=\bbC$ reduction}

In this case, the reduction data is 
\begin{equation}
   \tpsi(u) = u (r\del_r-\ii\xi) \ , 
   \qquad 
   \mathfrak{h} = 0\ ,
   \qquad 
   \mu = 0 \ . 
\end{equation}
Given $\mathfrak{h}$ is trivial, we have
\begin{equation}
   K = \{ u(r\del_r-\ii\xi) \} \ . 
\end{equation}
As before, we have already seen that $\tpsi$ is an isotropic lifted
action and so $(\tpsi,\mathfrak{h},\mu)$ are suitable reduction
data. $\mathcal{J}_-$ and $G$ are both invariant under $\tpsi(u)$ and
finally, using~\eqref{G-action}, we now have 
\begin{equation}
   GK = \{ u'(\dd\log r - \ii \eta) \} \ , 
\end{equation}
and hence
\begin{equation}
   E_K = \{ u(r\del_r-\ii\xi) + u'(\dd\log r - \ii \eta) \} \ , 
\end{equation}
Again we immediately see that
$\mathcal{J}_-E_K=E_K$. Hence, again assuming the action of $\xi_v$ on
$Y$ gives a $U(1)$ fibration, using the generalized Hermitian reduction
theorem,  we see that we have a generalized Hermitian 
structure on $\Ered$ over the four-dimensional space $\Mred = X/\bbC^*
= Y/U(1)$, or more generally, we get a generalized Hermitian structure
on the transversal section to the $r\del_r-\ii\xi_v$ orbits. 

Note that in both cases the reduced manifold $\Mred$ is the
same. Furthermore, the (complexified) spaces $E_K$, and hence the $G$-orthogonal
complements $K^G$, also agree. As discussed in~\cite{bcg}, $K^G$ is
a model for the reduced bundle $\Ered$. Thus the two reductions give
identical generalized Hermitian structures on $\Mred$.

%%%%%%%%%%%%%%%%%%%%%%%%%%%%%%%%%%%%%%%%%%%%%%%%%%%%%%%%%%%%%%%%%%%%%%%

\subsection{The reduced pure spinors}
\label{sec:Omega-red}

We now calculate the conditions on the reduced generalized Hermitian
structure implied by supersymmetry. The reduced structure can be
defined by a pair of commuting generalized almost complex structures:
$\LJ$ which is integrable and is the reduction of $\mathcal{J}_-$, and
a non-integrable structure $\RJ$, defined such that $-\LJ\RJ$ is the
reduced generalized metric. Equivalently, the structures are defined
as a pair of pure spinors $\Lred$ and $\Rred$. It is the differential
conditions on $\Lred$ and $\Rred$ implied by supersymmetry that we
will derive. 

In order to construct the reduced pure spinors, first note that the
reduction gives a splitting of the generalized tangent space 
\begin{equation}
\label{eq:Esplit}
   E = E_K \oplus K^G
\end{equation}
such that, in general, the $O(d,d)$ metric $\mukai{\cdot}{\cdot}$
factors into an $O(p,p)$ metric on $K^G$ and an $O(d-p,d-p)$ metric on
$E_K$. Thus we can similarly decompose sections of the spinor bundles
$S^\pm(E)$ into spinors of $\Spin(d-p,d-p)\times\Spin(p,p)$. In
particular, generic sections $\Omega_\pm\in S^\pm(E)$ can be written
as 
\begin{equation}
   \Omega_\pm = \Theta_\pm \otimes \tilde{\Omega}_+ 
      \oplus \Theta_\mp \otimes \tilde{\Omega}_- \ . 
\end{equation}
It is then the spinor components of $\tilde{\Omega}_\pm$ in $S^\pm(K_G)$ which
correspond to the reduced pure spinors. For the case in hand the
relevant decomposition is under
$\Spin(2,2)\times\Spin(4,4)\subset\Spin(6,6)$. As we will see below,
the reduction is such that the pure spinors defining the 
supersymmetric background decompose as 
\begin{equation}
\label{genspin-decomp}
\begin{aligned}
   \Omega_- = \Theta_- \otimes \Lred \ , \\
   \Omega_+ = \Theta_+ \otimes \Rred \ . 
\end{aligned}
\end{equation}
Thus the reduced spinors $\Lred$ and $\Rred$ are both positive
helicity in $\Spin(4,4)$. 

To make this explicit we need a basis for the $\Spin(6,6)$ gamma
matrices reflecting the decomposition~\eqref{eq:Esplit}. We first
introduce coordinates adapted to the reduction. We write the
$R$-symmetry Killing vector as\footnote{Note that this, more
  conventional, normalization of $\psi$ differs from the corresponding
  coordinate in~\cite{Gauntlett:2005ww} by a factor of three.}
\begin{equation}
\label{psi-def}
   \xi_v = K_5^\# = \del_\psi \ , 
\end{equation}
Let $y^m$ be coordinates on the transversal section to the
$R$-symmetry foliation. This means that $i_{\xi_v}\dd y^m=0$ and, in
particular, the metric decomposes as 
\begin{equation}
   g_Y = K_5\otimes K_5 + \gred_{mn}\dd y^m \dd y^n \ , 
\end{equation}
in analogy to~\eqref{eq:KEred}. The reduction structure already
defines a natural basis on $E_K$ given by 
\begin{equation}
\begin{aligned}
   \hat{f}_1 &= r\del_r \ , & 
      f^1 &= \dd \log r \ , \\
   \hat{f}_2 &= \xi \ , & 
      f^2 &= \eta \ , \\
\end{aligned}
\end{equation}
and satisfying $\mukai{f^i}{\hat{f}_j}=\frac{1}{2}\delta^i{}_j$ and
$\mukai{f^i}{f^j}=\mukai{\hat{f}_i}{\hat{f}_j}=0$. We can
then define an orthogonal basis on $K^G$ given by 
\begin{equation}
\begin{aligned}
   \hat{e}_m &= \me^{-b_2}\del_{y^m} - \tilde{\eta}_m\xi \ , & 
      e^m &= \dd y^m - \eta^m\xi
\end{aligned}
\end{equation}
where $\tilde{\eta}_m=2\mukai{\eta}{\me^{-b_2}\del_{y^m}}$ and 
$\eta^m=2\mukai{\eta}{\dd y^m}=i_{\eta_v}\dd y^m$. This 
basis again satisfies $\mukai{e^m}{\hat{e}_n}=\frac{1}{2}\delta^m{}_n$ and
$\mukai{e^m}{e^n}=\mukai{\hat{e}_m}{\hat{e}_n}=0$.

Given such a basis we can then write a generic $\Spin(6,6)$ spinor
using the standard raising and lowering operator
construction. Consider the polyform $\Omega^{(0)}=\me^{-b_2}\in
S^+(E)$. It is easy to see that we have the Clifford actions 
\begin{equation}
   \hat{f}_i \cdot \Omega^{(0)} = \hat{e}_m \cdot \Omega^{(0)} = 0 \ , 
\end{equation}
for all $i$ and $m$. Thus we can regard $\Omega^{(0)}$ as a ground
state for the lowering operators $(\hat{f}_i,\hat{e}_m)$. A generic
spinor is then given by acting with the anti-commuting raising
operators $(f^i,e^m)$. Acting with the $e^m$ first, we see that a
generic (non-chiral) spinor has the form
\begin{equation}
\label{Odecomp}
   \Omega = \me^{-b_2}\tilde\Omega_0 
      + f^1\cdot\me^{-b_2}\tilde\Omega_1 
      + f^2\cdot\me^{-b_2}\tilde\Omega_2 
      + f^1\cdot f^2\cdot\me^{-b_2}\tilde\Omega_3 \ , 
\end{equation}
where $\tilde{\Omega}_i$ are polyforms in $\dd y^m$, and
$\me^{-b_2}\tilde{\Omega}_i$ transform as a $\Spin(4,4)$ spinor under
the Clifford action of $(e^m,\hat{e}_m)$. 

We can now write the supersymmetry pure spinors $\Omega_\pm$
in the form~\eqref{Odecomp}. Requiring that $r\del_r-\ii\xi$ and
$\dd\log r-\ii\eta$ annihilate $\Omega_-$ while
$r\del_r-\ii\me^{2\Delta+\phi/2}\eta$ and $\dd\log
r-\ii\me^{-2\Delta-\phi/2}\xi$ annihilate $\Omega_+$ one finds the
only possibility is 
\begin{equation*}
\begin{aligned}
   \Omega_- &= \left( \dd\log r - \ii \eta\right)
       \cdot r^3\me^{-3\ii\psi}\me^{-b_2}\Lred \ , \\
   \Omega_+ &= \left(1 + \ii\me^{2\Delta+\phi/2} 
        \dd\log r\cdot \eta\right)\cdot r^3\me^{-b_2}\Rred \ , 
\end{aligned}
\end{equation*}
where $\Lred$ and $\Rred$ are both even polyforms in $\dd y^m$, as
claimed in~\eqref{genspin-decomp}. We have introduced factors of $r^3$
and $\me^{-3\ii\psi}$ so that $\Lred$ and $\Rred$ are independent of
the $r$ and $\psi$ coordinates. In general, they are then only locally
defined. 

One can then derive the conditions on $\Lred$ and $\Rred$, reduced to
the transversal section, implied by supersymmetry. From
$\dd\Omega_-=0$ one finds 
\begin{equation}
   \dd\Lred = - 3\ii \etar \cdot \Lred \ , 
\end{equation}
where $\etar$ is generalized vector on the transversal section defined
by 
\begin{equation}
   \eta = \dd\psi + \me^{-b_2}\etar \ . 
\end{equation}
This is means that $\Lred$ defines an integrable generalized complex
structure on the transverse section, as expected from the reduction
theorem. 

For the second pure spinor, the condition~\eqref{ReSUSY} on
$\re\Omega_+$ is equivalent to
\begin{equation}
\label{red1}
\begin{aligned}
   \dd\big( \me^{\Delta+\phi/4}\etar\cdot\im\Rred\big)
       &= - 2 \me^{-\Delta-\phi/4}\re\Rred \ , \\
    \dd\big( \me^{\Delta+\phi/4}\im\Rred\big) &= 0  \ . 
\end{aligned}
\end{equation}
Finally, since $i_{r\del_r}F=0$, following~\eqref{Odecomp}, we can
decompose the flux as  
\begin{equation}
   \me^{-B}F = \me^{-b_2}\tilde{F} + \eta\cdot\me^{-b_2}\tilde{G} . 
\end{equation}
The final condition~\eqref{djomp} is then equivalent to 
\begin{equation}
\label{redF}
\begin{aligned}
   \dd\big( 
      \me^{-\Delta-\phi/4}\etar\cdot\re\Rred \big) 
      &= -\tfrac{1}{8}\tilde{G}\ , \\
   \big[\tilde{\mathcal{J}_1}\,\dd(4\Delta+\phi)\big]\cdot\im\Rred
      &= -\tfrac{1}{8}\me^{3\Delta+3\phi/4}\tilde{F} \ , 
\end{aligned}
\end{equation}   
and to
\begin{equation}
\label{red2}
   \tilde{\mathcal{J}}_1\cdot\dd\big(
      \me^{-\Delta-\phi/4}\etar\cdot\re\Rred \big) = 0 \ , 
\end{equation}
where $\tilde{\mathcal{J}}_1$ is the reduction to the
transverse section of the generalized complex structure
$\mathcal{J}_-^{b_2}=\me^{b_2}\mathcal{J}_-\me^{-b_2}$, and we have
used the compatibility relation
$\tilde{\mathcal{J}}_1\cdot\Rred=0$.\footnote{Note that in the
  language defined in section~\ref{sec:BPS} below, the 
  condition~\eqref{red2} states that 
  $\dd\big(\me^{-\Delta-\phi/4}\etar\cdot\re\Rred\big)$ (and hence
  $\tilde{G}$) is an element of $U^0_{\tilde{\mathcal{J}}_1}$.} 

The conditions~\eqref{red1} and~\eqref{red2}, which do not involve the
flux, can be viewed as a generalization of the usual K\"ahler-Einstein
conditions. Given an $\Rred$ satisfying these conditions, the flux is
then determined by~\eqref{redF}.

%%%%%%%%%%%%%%%%%%%%%%%%%%%%%%%%%%%%%%%%%%%%%%%%%%%%%%%%%%%%%%%%%%%%%%%%%%%%%%%%%%%%%%
\section{The pure spinor $\Omega_{-}$}
\label{ompgenminus}
%%%%%%%%%%%%%%%%%%%%%%%%%%%%%%%%%%%%%%%%%%%%%%%%%%%%%%%%%%%%%%%%%%%%%%%%%%%%%%%%%%%%%%

The closed pure spinor $\Omega_-$ is associated with the integrable generalized complex-structure
${\cal J}_-$. The latter in turn holds 
information regarding BPS operators in the dual field theory. In this section we explore two aspects of this duality. The first 
is the mesonic moduli space of the dual theory, which is known to correspond to the subspace for which the polyform $\Omega_-$ 
 reduces to a three-form. The second is the connection between generalized holomorphic objects and dual BPS operators.

%%%%%%%%%%%%%%%%%%%%%%%%%%%%%%%%%%%%%%%%%%%%%%%%%%%%%%%%%%%%%%%%%%%%%%%%%%%%%%%%%%%%%%
\subsection{The general form of $\Omega_-$}
\label{genO-}
%%%%%%%%%%%%%%%%%%%%%%%%%%%%%%%%%%%%%%%%%%%%%%%%%%%%%%%%%%%%%%%%%%%%%%%%%%%%%%%%%%%%%%

Recall that the most general pure spinor takes the form \cite{gualtieri}
\be
\Omega = \alpha \theta_1 \wedge \theta_2 \wedge ... \wedge \theta_k\wedge \me^{-b+\ii\omega^0}~,
\ee
where $\alpha$ is some complex function, $\theta_i$ are complex one-forms, while $b$ and $\omega^0$ are both real two-forms. The integer $k$ is called the \emph{type} of the pure spinor, which can change along various subspaces of $X$.

Using the definition of $\Omega_\pm$, the Fierz identity \reef{eq:Fierz}, and the
results of section \ref{sec:AdS} and appendix \ref{sec:56}, one can find expressions for $\Omega_\pm$ in terms of spinor bilinears introduced in
\cite{Gauntlett:2005ww}.
We find that in general $\Omega_-$ is of type one with
\be
\label{type1}
\Omega_- = \theta \wedge \me^{-b_- + \ii \omega_-} \;,
\ee
where
\bea
\theta&=&-\frac{r^3}{8}\me^{4\Delta}(\ii K+S\dd\log r)~,\nn
\omega_- &=&\frac{4\me^{6\Delta+\phi/2}}{f(\sin2\Bphi)^2} \left( K_5\wedge \im(K_3)-\cos2\Btheta\cos2\Bphi\re(K_3)\wedge \dd \log r \right)~, \nn
b_- &=& -\frac{4\me^{6\Delta+\phi/2}}{f(\sin2\Bphi)^2} \left( K_4\wedge \re(K_3)+(\cos2\Bphi)^2\im(K_3)\wedge \dd \log r\right) + b_2~.
\eea
Note that $\omega_-$ and $b_-$ are not uniquely defined since we can add two-forms that vanish when wedged with $\theta$. 
Here the angles $\Btheta$ and $\Bphi$, which appear in appendix B of \cite{Gauntlett:2005ww} without bars, 
are functions on the link $Y$ that are related to the scalar spinor bilinears through
\bea\label{tspb}
\sin \zeta &=& \cos2\Btheta \cos2\Bphi  \;, \\
|S| &=& - \sin2\Btheta \cos2\Bphi \;.
\eea
Using the results of \cite{Gauntlett:2005ww}, we have the important
result that $\theta$ is  exact\footnote{The fact that $\theta$ is closed was essentially observed
in~\cite{Martucci:2006ij}, and it was also shown to be exact in the
special cases of the Pilch-Warner and Lunin-Maldacena solutions
in~\cite{Minasian:2006hv}.}
\be
\theta=\dd\left[-\frac{1}{24}\me^{4\Delta} r^3S\right] \equiv \dd \left( r^3 \theta_0 \right)\;. \label{exacttheta}
\ee
Alternatively, from the supersymmetry equation
$\diff\Omega_-=0$ and the definite scaling dimension ${\cal L}_{r\partial_r} \Omega_- = 3\Omega_-$,
we immediately obtain
\bea
\Omega_- &=& \tfrac{1}{3} \dd (r\partial_r \lrcorner\Omega_-)~,
\eea
the one-form part of which reduces to \eqref{exacttheta}. 

%%%%%%%%%%%%%%%%%%%%%%%%%%%%%%%%%%%%%%%%%%%%%%%%%%%%%%%%%%%%%%%%%%%%%%%%%%%%%%%%%%%%%%
\subsection{Type change of $\Omega_-$ and the mesonic moduli space}\label{sec:type}
%%%%%%%%%%%%%%%%%%%%%%%%%%%%%%%%%%%%%%%%%%%%%%%%%%%%%%%%%%%%%%%%%%%%%%%%%%%%%%%%%%%%%%

The pure spinor $\Omega_-$ has the property that its type can jump from type one to type three on the locus $\theta=0$.
This locus can be neatly parameterized through the angles $\Btheta$ and $\Bphi$. 
Assuming $f\ne 0$, we have from \reef{fxf} that $\sin\zeta$ is nowhere zero and then \reef{tspb}
implies that both $\cos2\Bphi$ and $\cos2\Btheta$ are nowhere zero. 
 Using the expression for $K$ in appendix B of \cite{Gauntlett:2005ww}, we see that when $f\ne 0$
\bea
\sin2\Btheta = 0 &\Longleftrightarrow& \theta_0 = 0 \;, \\
\sin2\Btheta = \sin2\Bphi = 0 &\Longleftrightarrow& \theta = 0 \;.
\eea
The locus $\theta=0$ is thus a sublocus of $\theta_0=0$.
Notice that, where $\theta=0$, $\Omega_-$ is not identically zero, as
one might have naively expected  from~\eqref{type1}, but instead
reduces to a finite, non-zero three-form. Indeed, the powers of
$\sin2\Bphi$ in the denominator of $b_-$ and $\omega_-$ are cancelled
by those in $K$, $K_3$ and $K_4$. 

The locus $\theta=0$ is precisely where a probe pointlike D3-brane in $X$ is supersymmetric. This follows from
\cite{Martucci:2006ij} where it was shown that the pull-back of $\theta$ to the D3-brane worldvolume is equal to the
F-term of the worldvolume theory. The supersymmetric locus of such a pointlike
D3-brane is naturally interpreted as the mesonic moduli space. 

%%%%%%%%%%%%%%%%%%%%%%%%%%%%%%%%%%%%%%%%%%%%%%%%%%%%%%%%%%%%%%%%%%%%%%%%%%%%%%%%%%%%%%
\subsection{BPS operators and generalized holomorphic spinors}
\label{sec:BPS}
%%%%%%%%%%%%%%%%%%%%%%%%%%%%%%%%%%%%%%%%%%%%%%%%%%%%%%%%%%%%%%%%%%%%%%%%%%%%%%%%%%%%%%
In the Sasaki-Einstein case, holomorphic functions on the Calabi-Yau cone with a definite scaling weight $\lambda$ under the action of $r\del_r$ also have a charge $\lambda$ under the action of $\xi$. This 
stems from the intimate connection (via Kaluza-Klein reduction on the Sasaki-Einstein manifold)
between holomorphic functions on the cone and BPS operators in the dual CFT,
in fact (anti-)chiral primary operators.
For general $AdS_5$ solutions we might expect that the holomorphic functions should be replaced by polyforms and that 
the BPS condition of matching charges should be with respect to the generalized Lie derivative $\Lgen$ discussed in section 2. 
We now derive such a result, leaving the detailed connection with 
Kaluza-Klein reduction on the internal space $Y$ to future work.

We first recall that a generalized almost complex structure $\mathcal{J}$ defines a
grading on generalized spinors, or equivalently differential forms. If $\Omega\in \SC {S_\pm(E)}$ is a pure spinor corresponding to $\mathcal{J}$, one
defines the canonical pure spinor line bundle $U_{\cal J}^n\subset
S_\pm(E)$ as sections of the form $\varphi=f\Omega$ for some function
$f$. One can then define 
\be
U_{\cal J}^{(n-k)} = \wedge^k \bar{L}\otimes \, U_{\cal J}^n \;.
\ee
Elements of $U_{\cal J}^k$ have eigenvalues $\ii k$ under the Lie
algebra action of ${\cal J}$ given in~\eqref{eq:Jaction}. These
bundles then give a grading of the spinor bundles $S_\pm(E)$. 
A generalized vector $V \in \Gamma(E)$ acting on an element of $U_{\cal J}^k$ gives an element of $U_{\cal J}^{k+1}\oplus U_{\cal J}^{k-1}$. In particular an annihilator of $\Omega$ acts by purely raising the level by one.
If the generalized complex structure ${\cal J}$ is also integrable then the exterior derivative splits into the sum
\be
\dd = \del_{\cal J} + \delb_{\cal J} \;,
\ee
where
\be\label{delbar}
C^{\infty}\left(U_{\cal J}^k\right) \begin{array}{c} \underleftarrow{\delb_{\cal J}} \\ \overrightarrow{\del_{\cal J}}  \end{array} C^{\infty}\left(U_{\cal J}^{k-1}\right) \;.
\ee

Consider now a spinor $\psi$ satisfying
\bea
\psi &\in& U^k_{{\cal J}_-} \;, \nonumber \\
\Lgen_{r\del_r} \psi &=& \lambda \psi \;, \label{holconst}
\eea
for some $k$ and $\lambda$. Then imposing in addition 
\bea
\bar\del_{{\cal J}_-}\psi=0~, \qquad (r\del_r+\ii\xi)\cdot\psi=0\qquad \mbox{implies} \qquad\Lgen_\xi\psi=\ii\Lgen_{r\del_r}\psi~.
\eea
In other words, subject to the constraints (\ref{holconst}), a spinor is BPS if it is generalized holomorphic and is annihilated by $r\del_r +\ii \xi$. 
To see this result, we first write $r\del_r=(1/2)(r\del_r+\ii\xi)+(1/2)(r\del_r-\ii\xi)$ and use
\reef{holconst} to deduce that
\bea
\del_{{\cal J}_-}[(r\del_r+\ii\xi)\cdot\psi]+(r\del_r+\ii\xi)\cdot\del_{{\cal J}_-}\psi=0~,\nn
\bar\del_{{\cal J}_-}[(\del_r-\ii\xi)\cdot\psi]+(r\del_r-\ii\xi)\cdot\bar\del_{{\cal J}_-}\psi=0~.
\eea
In obtaining this we used the fact that since $ r\del_r - \ii \xi$ is an annihilator of $\Omega_-$ it raises the level of $\psi$ and similarly
$ r\del_r + \ii \xi$ lowers the level.
We then compute
\bea
\Lgen_{\xi} \psi &=& \ii \Lgen_{r\del_r} \psi - \ii\left\{ \dd \left[ \left( r\del_r + \ii \xi \right)\cdot \psi \right] +  \left( r\del_r + \ii \xi \right)\cdot\dd\psi \right\}  \nn
&=& \ii \Lgen_{r\del_r} \psi - \ii\left\{ \bar{\del}_{{\cal J}_-} \left[  \left( r\del_r + \ii \xi \right)\cdot  \psi \right] +  \left( r\del_r + \ii \xi \right)\cdot \bar{\del}_{{\cal J}_-}\psi \right\} \;. \label{BPSliehol}
\eea
In a similar way, given \reef{holconst} we also have
\be
\del_{{\cal J}_-}\psi=0~,\qquad (r\del_r-\ii\xi)\cdot\psi=0\qquad \mbox{implies} \qquad\Lgen_\xi\psi=-\ii\Lgen_{r\del_r}\psi~.
\ee

%%%%%%%%%%%%%%%%%%%%%%%%%%%%%%%%%%%%%%%%%%%%%%%%%%%%%%%%%%%%%%%%%%%%%%%%%%%%%%%%%%%%%%%
\section{The pure spinor $\Omega_+$}
\label{ompgenplus}
%%%%%%%%%%%%%%%%%%%%%%%%%%%%%%%%%%%%%%%%%%%%%%%%%%%%%%%%%%%%%%%%%%%%%%%%%%%%%%%%%%%%%%%
\subsection{The general form of $\Omega_+$}
\label{genO+}
One can see immediately from the supersymmetry equation \reef{ImSUSY}
that if we assume $F_5 \neq 0$, which we shall do,
then $\im{\Omega_+}$ must have a scalar component and hence
$\Omega_+$ is of type $0$:
\be
\Omega_+ = \alpha_+ \me^{-b_+ + \ii\omega_+^0}~. \label{genppans}
\ee
Using the same procedure as in the last section, we may again express these
quantities in terms of the bilinears of
\cite{Gauntlett:2005ww}. After defining the rescaled two-form
\be\label{omegadef}
\omega=\me^{-2A}r^4\omega^0_+~,
\ee
we find
\bea\label{tranres}
\alpha_+&=&-\frac{\ii}{32}f \me^{-A}r^4~,\nn
\omega &=& -\frac{4r^2}{f} \me^{4\Delta} \left( V + K_4 \wedge \dd \log r \right)~,\nn
b_+ &=& \me^{6\Delta + \phi/2} \frac{4}{f} \im{K_3}\wedge \dd\log r+b_2~,
\eea
where $b_2$ appears in \reef{gaugeb}.
\subsection{A canonical symplectic structure}
The rescaling (\ref{omegadef}) is motivated by the fact that $\omega$ defines a canonical symplectic structure.
To see this, we first observe that $Y$ admits a contact structure defined by the one-form
\be\label{sigmadef}
\sigma \equiv \frac{4}{f} \me^{4\Delta} K_4~.
%\nonumber
\ee
Recall that for a one-form $\sigma$ to be contact, the top-degree form $\sigma\wedge \dd\sigma \wedge \dd\sigma$
must be nowhere vanishing. Using (3.19) of \cite{Gauntlett:2005ww}, and results in appendix B
of \cite{Gauntlett:2005ww}, one can easily show that 
\be\label{volform}
\sigma \wedge \dd\sigma \wedge \dd\sigma
   = \frac{128}{f^2} \me^{8\Delta} \widetilde{\vol}_Y
   = \frac{8}{\sin^2\zeta}\widetilde{\vol}_Y
~,
\ee
where recall $\widetilde{\vol}_Y=-e^{12345}$ (using the orthonormal frame in
appendix B of \cite{Gauntlett:2005ww}). 
We then observe, using (3.19) of \cite{Gauntlett:2005ww}, that
\bea\label{ppan2}
\omega = \tfrac{1}{2}\dd ( r^2 \sigma)~,
%\nonumber
\eea
which shows that $\omega$ is closed and non-degenerate, and hence defines a symplectic
structure on the cone $X=\R^+\times Y$. Alternatively, one can see the formula
(\ref{ppan2}) for $\omega$ directly from the supersymmetry equation (\ref{ReSUSY})
on noting that $\me^{-A}\Omega_+$ has scaling dimension 2 under $r\del_r$.
Furthermore, again using the results of appendix B of~\cite{Gauntlett:2005ww}, we
have
\bea\label{Reeb}
1=\xi_v\lrcorner \sigma~,\qquad
0=\xi_v\lrcorner\diff\sigma~,
\eea
which shows that $\xi_v$ is also the unique ``Reeb vector field''
associated with the contact structure.
Notice also that (\ref{ppan2}) implies
that $\mathcal{H}=r^2/2$ is precisely the Hamiltonian function for the Hamiltonian
vector field $\xi_v$, {\it i.e.} $\diff \mathcal{H} = -i_{\xi_v}\omega$.
It is remarkable that these features, which are well-known in the Sasaki-Einstein case, are valid
for all supersymmetric $AdS_5$ solutions with non-vanishing five-form flux.

Although we have a symplectic structure, we do not quite
have a K\"ahler structure, as in the Calabi-Yau case, but it is quite close.
Using the last equation in (\ref{tranresvects}) and the definition (\ref{sigmadef}) we see that
\bea
\eta_f = \sigma + i_{\eta_v}b_2~,
\eea
and thus $\left(\me^{b_2}\eta\right)\mid_{1-\mathrm{form}} = \sigma$.
Since $\me^{b_2}(\diff \log r)=\diff \log r$ manifestly, and by
definition $\eta\equiv\mathcal{J}_-(\diff\log r)$, we have, using \reef{eq:componentJ},
\be
\sigma = \mathcal{J}_-^{b_2} (\diff \log r)\mid_{1-\mathrm{form}} = -(I^{b_2}_-)^*(\diff\log r)~.
\ee
Note this is precisely analogous to the formula for the contact form in the Sasakian case.
We then have
\bea
\diff^{\mathcal{J}_-^{b_2}} r^2 = -r^2 \, \diff \left(Q^{b_2}_- +\frac{1}{2}\mathrm{Tr}\, I^{b_2}_-\right)     - (I^{b_2}_-)^*(\diff (r^2))~,
\eea
where here we recall that in general we define $\dd^{\cal J_-}\equiv -[\dd,{\cal J}_-]$, and we use
\reef{eq:Jaction} for the action on generalized spinors. From 
this it follows that
\bea \label{kaehlerpotential}
\omega = \frac{1}{4}\diff\diff^{\mathcal{J}_-^{b_2}} r^2 +\frac{1}{4}\diff(r^2)\wedge \diff \left(Q^{b_2}_- +\frac{1}{2}\mathrm{Tr}\, I^{b_2}_-\right)~.
\eea
Thus $r^2$ is almost a K\"ahler potential, for the $b_2$-transformed complex structure
$\mathcal{J}_-^{b_2}=\me^{b_2}\mathcal{J}_-\me^{-{b_2}}$, except for the last term. 

%%%%%%%%%%%%%%%%%%%%%%%%%%%%%%%%%%%%%%%%%%%%%%%%%%%%%%%%%%%%%%%%%%%%%%%%%%%%%%%%%%%%%%
\subsection{The central charge as a Duistermaat-Heckman integral}
%%%%%%%%%%%%%%%%%%%%%%%%%%%%%%%%%%%%%%%%%%%%%%%%%%%%%%%%%%%%%%%%%%%%%%%%%%%%%%%%%%%%%%
Recall that in any four-dimensional CFT there are two central charges, usually
called $a$ and $c$, that are constant coefficients in the conformal anomaly
\bea
\langle T_\mu^\mu \rangle = \frac{1}{120 (4\pi)^2}\left(c (\mathrm{Weyl})^2 - \frac{a}{4}(\mathrm{Euler})\right)~.
\eea
Here $T_{\mu\nu}$ denotes the stress-energy tensor, and Weyl and Euler denote
certain curvature invariants for the background four-dimensional metric.
For SCFTs, both $a$ and $c$ are related to the $R$-symmetry \cite{Anselmi:1997ys} via
\bea
a=\frac{3}{32}\left(3\mathrm{Tr}R^3 - \mathrm{Tr}R\right), \qquad
c=\frac{1}{32}\left(9\mathrm{Tr}R^3 - 5\mathrm{Tr}R\right)~.
\eea
Here the trace is over the fermions in the theory.
For SCFTs with $AdS_5$ gravity duals, in fact $a=c$ holds necessarily 
in the large $N$ limit \cite{Henningson:1998gx}.
The central charge of the SCFT is then inversely proportional to the dual five-dimensional
Newton constant $G_5$ \cite{Henningson:1998gx}, obtained here by Kaluza-Klein
reduction on $Y$. The Newton constant, in turn, was computed in
appendix E of \cite{Gauntlett:2005ww}, and is given by
\bea\label{5dNewton}
G_5 = \frac{G_{10}}{V_5} = \frac{\kappa^2_{10}}{8\pi V_5}~,
\eea
where $G_{10}$ is the ten-dimensional Newton constant of
type IIB supergravity, and we have defined
\bea
V_5 \equiv \int_Y \me^{8\Delta}\widetilde{\vol}_Y~.
\eea

We may derive an alternative formula for $G_5$ as follows. We begin by
rewriting
\bea
V_5 = \frac{f^2}{16}\int_Y \frac{1}{\sin^2\zeta}\widetilde{\vol}_Y~,
\eea
where we have used the relation (\ref{fxf}).
Importantly, the constant $f$ is quantized, being essentially the number of
D3-branes $N$. Specifically, we have
\bea
N =\frac{1}{(2\pi l_s)^4g_s}\int_Y \dd C_4=
\frac{1}{(2\pi l_s)^4g_s}
\int_Y \left(F_5 + H \wedge C_2\right)~.
\eea
Using the Bianchi identity $DG=-P\wedge G^*$ and the result \reef{ddub},
one derives that $\dd(H\wedge C_2)=-(2/f)\dd[\me^{6\Delta}\mathrm{Im}(W^*\wedge G)]$ and so
we can also write
\be
N= \frac{1}{(2\pi l_s)^4g_s}\int_Y \left(F_5 - \frac{2\me^{6\Delta}}{f} \im{\left[ {W}^* \wedge G \right]}\right)~.
\ee
We may evaluate this expression in terms of the orthonormal basis of forms $e^i$ introduced in
appendix B of \cite{Gauntlett:2005ww}, and after some calculation we find
\bea\label{Nf}
N= -\frac{f}{(2\pi l_s)^4g_s}\int_Y \frac{1}{\sin^2 \zeta} \, \widetilde{\vol}_Y~.
\eea
Combining these formulae and using
$2\kappa_{10}^2=(2\pi)^7l_s^8g_s^2$ leads to the result
\bea
G_5 =\frac{8V_5}{\pi^2f^2N^2}~.
\eea

Consider now the integral
\bea\label{muconstant}
\mu = \frac{1}{(2\pi)^3}\int_X \me^{-r^2/2}\frac{\omega^3}{3!}~.
\eea
This is the Duistermaat-Heckman integral for a symplectic manifold
$(X,\omega)$ with Hamiltonian function $\mathcal{H}=r^2/2$, which we
have shown is the Hamiltonian for the Reeb vector field $\xi_v$.
Using \reef{ppan2} and \reef{volform} we may rewrite
\bea
\frac{\omega^3}{3!} =
\frac{16}{f^2} \me^{8\Delta}r^5\diff r\wedge\widetilde{\vol}_Y~.
\eea
Performing the $r$-integral in (\ref{muconstant}) allows us to rewrite the five-dimensional Newton constant as
\bea
G_5 = \frac{\pi\mu}{2N^2}~.
\eea
Since $\mu=1$ for the round five-sphere solution, we thus obtain the ratio
$\frac{G_5}{G_{S^5}} = \mu$. Recalling that this is, by AdS/CFT duality, the inverse ratio of central charges \cite{Henningson:1998gx},
we deduce the key result
\bea\label{centralcharge}
\frac{a_{\mathcal{N}=4}}{a} = \frac{1}{(2\pi)^3}\int_X \me^{-r^2/2}\frac{\omega^3}{3!} =
\frac{1}{(2\pi)^3} \int_Y \sigma\wedge\diff\sigma\wedge\diff\sigma~.
\eea
Here $a_{\mathcal{N}=4} = {N^2}/{4}$ denotes the (large $N$) central charge
of $\mathcal{N}=4$ super-Yang-Mills theory.

The formula (\ref{centralcharge})
implies that the central charge depends only on the symplectic structure of the cone
$(X,\omega)$ and the Reeb vector field $\xi_v$. This is perhaps surprising: one might have anticipated that 
the quantum numbers of quantized fluxes would appear explicitly in the central charge formula. 
However, recall from formulae (\ref{gaugeb}), (\ref{gaugec}) that the two-form potentials 
$B$ and $C_2$ are globally defined. In particular, for example, the period of $H=\diff B$ through any three-cycle in $Y$ is zero.

As discussed in \cite{Martelli:2006yb},
the Duistermaat-Heckman integral in (\ref{centralcharge}) may be evaluated by localization.
The integral localizes where $\xi_v=0$, which is formally at the tip of the cone
$r=0$. Unless the differentiable and symplectic structure is smooth here (which is only
the case when $X\cup \{r=0\}$ is diffeomorphic to
$\R^6$), one needs
to equivariantly resolve the singularity in order to apply the localization
formula. Notice here that since $\xi_v$ preserves all the structure
on the compact manifold $(Y,g_Y,\sigma)$, the closure of its orbits defines a $U(1)^s$ action preserving all the structure, for some $s\geq 1$. Here we have used the fact that the isometry group of a compact Riemannian manifold is compact.
Thus $(X,\omega)$ comes equipped with a $U(1)^s$ action.

Rather than attempt to describe this in general, we focus here on the special
case where the solution is \emph{toric}: that is, there is a $U(1)^3$ action on
$Y$ under which $\sigma$, and hence $\omega$ under the lift to $X$,
is invariant. Notice that we do not necessarily require that the \emph{full} supergravity 
solution is invariant under $U(1)^3$ -- we shall illustrate this in the 
next section with the Pilch-Warner solution, where $\sigma$ and the metric 
are invariant under $U(1)^3$, but the $G$-flux is invariant only under 
a $U(1)^2$ subgroup. For the arguments that follow, it is only $\sigma$
(and hence $\omega$) that we need to be invariant under a maximal 
dimension torus $U(1)^3$.
In fact any such symplectic toric
cone is also an affine toric variety. This implies that
there is a (compatible) complex structure on $X$, and that
the $U(1)^3$ action complexifies to a holomorphic $(\C^*)^3$ action
with a dense open orbit. There is then always a
symplectic toric resolution $(X',\omega')$ of $(X,\omega)$, obtained by
toric blow-up. In physics language, this is because
one can realize $(X,\omega)$ as a gauged linear sigma model,
and one obtains $(X',\omega')$ by simply turning on generic
Fayet-Iliopoulos parameters.
One can also describe this in terms
of moment maps as follows.
The image of a symplectic toric cone under the moment map $\mu:X\rightarrow {\mathbb R}^3$ is a
strictly convex rational polyhedral cone (see \cite{Martelli:2006yb}).
Choosing a toric resolution $(X',\omega')$ then amounts to choosing any
simplicial resolution $\mathcal{P}$ of this polyhedral cone. Here $\mathcal{P}$ is the
image of $\mu':X'\rightarrow \R^3$.
Assuming the fixed points of  $\xi_v$ are all isolated, the localization
formula is then simply \cite{Martelli:2006yb}
\bea\label{DHformula}
\frac{1}{(2\pi)^3}\int_X \me^{-r^2/2}\, \frac{\omega^3}{3!} =
\sum_{\mathrm{vertices\ }\mathrm{p}\in{\cal P}}  \prod_{i=1}^3 \frac{1}{\langle
\xi_v, u^{\mathrm p}_i\rangle}~.
\eea
Here $u_i^{\mathrm p}$, $i=1,2,3$, are the three edge vectors of the
moment polytope $\mathcal{P}$ at the vertex point $\mathrm{p}$, and
$\langle \cdot, \cdot\rangle$ denotes the standard Euclidean
inner product on $\R^3$ (where we regard $\xi_v$ as being an element of the Lie
algebra $\R^3$ of $U(1)^3$). The
vertices of $\mathcal{P}$ precisely
correspond to the $U(1)^3$ fixed points of the symplectic toric resolution $X'={X}_{{\mathcal P}}$ of $X$.
Thus, remarkably, these results of \cite{Martelli:2006yb} hold in general,
even when there are non-trivial fluxes turned on and $X$ is not
Calabi-Yau.

%%%%%%%%%%%%%%%%%%%%%%%%%%%%%%%%%%%%%%%%%%%%%%%%%%%%%%%%%%%%%%%%%%%%%%%%%%%%%%%%%%%%%%
\subsection{The conformal dimensions of BPS branes}
%%%%%%%%%%%%%%%%%%%%%%%%%%%%%%%%%%%%%%%%%%%%%%%%%%%%%%%%%%%%%%%%%%%%%%%%%%%%%%%%%%%%%%
A supersymmetric D3-brane wrapped on $\Sigma_3\subset Y$
gives rise to a BPS particle in $AdS_5$. The quantum field $\Phi$
whose excitations give rise to this particle state then
couples, in the usual way in AdS/CFT, to a dual
chiral primary operator $\mathcal{O}=\mathcal{O}_{\Sigma_3}$ in the
boundary SCFT. More precisely, there is an asymptotic
expansion of $\Phi$ near the $AdS_5$ boundary
\bea\label{expand}
\Phi \sim \Phi_0 r^{\Delta-4} + A_{\Phi} r^{-\Delta}~,
\eea
where $\Phi_0$ acts as the source for $\mathcal{O}$
and $\Delta=\Delta(\mathcal{O})$ is the conformal dimension of $\mathcal{O}$.
In \cite{Martelli:2007mk}, following \cite{Klebanov:2007us},
it was argued that the vacuum expectation value
$A_{\Phi}$ of  $\mathcal{O}$ in a given asymptotically
$AdS_5$ background may be computed from $\me^{-S_E}$, where $S_E$ is the on-shell
Euclidean action of the D3-brane wrapped on $\Sigma_4=\R^+\times\Sigma_3$, where $\R^+$ is the $r$-direction.
In particular, via the second term in (\ref{expand}) this identifies the conformal dimension
$\Delta=\Delta(\mathcal{O}_{\Sigma_3})$
with the coefficient of the
logarithmically divergent part of the on-shell Euclidean action
of the D3-brane wrapped on $\Sigma_4$.
We refer to section 2.3 of \cite{Martelli:2007mk} for further details.

We are thus interested in the on-shell Euclidean action of a
supersymmetric D3-brane wrapped on $\Sigma_4=\R^+\times\Sigma_3$.
The condition of supersymmetry is equivalent to a generalized calibration
condition, namely
equation (3.16) of \cite{Martucci:2005ht}.
In our notation and conventions, this calibration condition reads
\bea\label{MSformula}
\mathrm{Re}\left[-\ii \Phi_+\wedge\me^{\mathcal{F}}\right]\mid_{\Sigma_4} =
\frac{|a|^2}{8}\sqrt{\det(h+\mathcal{F})}\, \diff x_1\wedge\cdots\wedge
\diff x_4~.
\eea
Here $h$ is the induced (string frame) metric on $\Sigma_4$,
and $\mathcal{F}=F-B$ is the gauge-invariant worldvolume gauge field,
satisfying
\bea
\diff \mathcal{F}=-H\mid_{\Sigma_4}~.
\eea
Recalling from section 3.2 that $|a|^2=\me^A$, we may then substitute for $\Phi_+$ in
terms of $\Omega_+$ using (\ref{Omegaplus}) and (\ref{genppans}) to obtain
\bea\label{caleqn}
\mathrm{Re}\left[-\ii \Phi_+\wedge\me^{\mathcal{F}}\right]\mid_{\Sigma_4} = \frac{f}{64}
\me^{A+\phi}\diff \log r\wedge \sigma \wedge \dd \sigma\mid_{\Sigma_4} -
\frac{f}{64}\me^{-3A+\phi}r^4 (F-b_+)^2\mid_{\Sigma_4}~,
\eea
where, as in \reef{tranres},
\bea
b_+ = \me^{6\Delta + \phi/2} \frac{4}{f} \im{K_3}\wedge \dd\log r+b_2~.
\eea
Here $b_2$ is a closed two-form, whose gauge-invariant information
is contained in its cohomology class in $H^2(X,\R)/H^2(X,\Z)$.
In writing $b_+$ in (\ref{caleqn})
we have chosen a particular representative two-form for the class
of $b_2$ in  $H^2(X,\R)/H^2(X,\Z)$. Then under any gauge transformation
of $b_+$ (induced from a $B$-transform of $\Omega_+$),
the worldvolume gauge field $F$ transforms by precisely the
opposite gauge transformation restricted to $\Sigma_4$, so that
the quantity $F-b_+$ is gauge invariant on $\Sigma_4$.
We now choose the worldvolume
gauge field $F$ to be
\bea\label{gaugechoice}
F=b_2\mid_{\Sigma_4}~,
\eea
so that (\ref{caleqn}) becomes simply
\bea
\mathrm{Re}\left[-\ii \Phi_+\wedge\me^{\mathcal{F}}\right]\mid_{\Sigma_4} = \frac{f}{64}
\me^{A+\phi}\diff \log r\wedge \sigma \wedge \dd \sigma\mid_{\Sigma_4}~.
\eea
In fact, there is a slight subtlety in (\ref{gaugechoice}).
If the cohomology class of $b_2/(2\pi l_s)^2\mid_{\Sigma_4}$ in $H^2(\Sigma_4,\R)$
is not integral, then the choice (\ref{gaugechoice}) is not possible as $F$ is the curvature
of a unitary line bundle. Having said this, notice
$H^2(\Sigma_4,\R)\cong H^2(\Sigma_3,\R)$, and
thus in particular that
if $H^2(\Sigma_3,\R)=0$ then every closed $b_2\mid_{\Sigma_4}$ is exact,
and thus may be gauge transformed to zero on $\Sigma_4$. Then
(\ref{gaugechoice}) simply sets $F=0$. For every example of
a supersymmetric $\Sigma_3$ that we are aware of, this is indeed the case.
In any case, we shall assume henceforth that the choice (\ref{gaugechoice}) is
possible.

The calibration condition (\ref{MSformula}) for a D3-brane with worldvolume
$\Sigma_4$ and with gauge field (\ref{gaugechoice}) is thus
\bea\label{calDBI}
\frac{f}{8}\diff \log r\wedge\sigma\wedge \dd \sigma =
\me^{-\phi}\sqrt{\det(h-B)}\, \diff x_1\wedge\cdots\wedge\diff x_4~.
\eea
Notice the right hand side is precisely the Dirac-Born-Infeld Lagrangian, up
to the D3-brane tension $\tau_3=1/(2\pi)^3 l_s^4 g_s$.
From (\ref{calDBI}), and the comments above on the scaling dimension
$\Delta(\mathcal{O}(\Sigma_3))$ of the dual operator $\mathcal{O}(\Sigma_3)$, we thus deduce
\bea
\Delta(\mathcal{O}(\Sigma_3)) = -\frac{\tau_3 f}{8}\int_{\Sigma_3}\sigma\wedge\diff\sigma~.
\eea
(The sign just arising from a convenient choice of orientation.) 
Using \reef{Nf} and \reef{volform} we have
\bea
f = -\frac{8(2\pi l_s)^4 g_s N}{\int_Y \sigma\wedge \dd\sigma\wedge \dd\sigma}~,
\eea
and hence
\bea\label{conformal}
\Delta(\mathcal{O}(\Sigma_3))
&=& \frac{2\pi N\int_{\Sigma_3}\sigma\wedge\dd \sigma}{\int_Y\sigma\wedge
\dd\sigma\wedge\dd\sigma}~.
\eea
This is our final formula for the conformal dimension of
the chiral primary operator dual to a BPS D3-brane
wrapped on $\Sigma_3$.
Since we may write
\bea\label{3to4}
\int_{\Sigma_3} \sigma\wedge\dd \sigma = \int_{\Sigma_4}
\me^{-r^2/2} \frac{\omega^2}{2!}~,
\eea
we again see that it depends only on the symplectic structure of $(X,\omega)$
and the Reeb vector field $\xi_v$.
This again may be evaluated by localization, having
appropriately resolved the tip of the cone $\Sigma_4$.

%%%%%%%%%%%%%%%%%%%%%%%%%%%%%%%%%%%%%%%%%%
\section{The Pilch-Warner solution}
%%%%%%%%%%%%%%%%%%%%%%%%%%%%%%%%%%%%%%%%%%

In this section we illustrate the general results derived so far with the 
Pilch-Warner solution \cite{Pilch:2000ej}. (Some aspects of the
generalized complex geometry of this background have already been
discussed in~\cite{Minasian:2006hv}.) Recall that the Pilch-Warner solution is dual to a
Leigh-Strassler fixed point theory \cite{Leigh:1995ep} which is
obtained by giving a mass to one of the three chiral superfields (in
$\mathcal{N}=1$ language) of $\mathcal{N}=4$ $SU(N)$ super-Yang-Mills
theory, and following the resulting renormalization group flow to the
IR fixed point theory. This latter theory is an ${\cal N}=1$ $SU(N)$
gauge theory with two adjoint fields $Z_a$, $a=1,2$, which form a
doublet under an $SU(2)$ flavour symmetry, and a quartic
superpotential. Since the superpotential has scaling dimension three,
this fixes $\Delta(Z_a)=3/4$, implying that the IR theory is strongly
coupled. The mesonic moduli space is simply $\mathrm{Sym}^N \C^2$.

The Pilch-Warner supergravity solution \cite{Pilch:2000ej} was rederived in
\cite{Gauntlett:2005ww}, and we shall use some of the results from that reference also.
We have $Y=S^5$ with non-trivial metric
\bea
g_Y &=& \frac{1}{9}\bigg[6\dd \vartheta^2 + \frac{6\cos^2\vartheta}{3-\cos2\vartheta}(\sigma_1^2+\sigma_2^2)
 + \frac{6\sin^22\vartheta}{(3-\cos2\vartheta)^2}\sigma_3^2 \nn &&\qquad\qquad\qquad\qquad\qquad+4\left(\dd\varphi+\frac{2\cos^2\vartheta}{3-\cos2\vartheta}\sigma_3\right)^2\bigg]~,
\eea
where $0\leq\vartheta\leq\tfrac{\pi}{2}$, $0\leq \varphi\leq 2\pi$, and $\sigma_i$, $i=1,2,3$, are left-invariant
one-forms on $SU(2)$
(denoted with hats in \cite{Gauntlett:2005ww}). The dilaton $\phi$ and axion
$C_0$ are simply constant, while the warp factor is
\bea
\me^{4\Delta} = {\frac{f}{8}(3-\cos2\vartheta)}~.
\eea
There is also a non-trivial NS and RR three-form flux given by (see \reef{psandqs})
\bea\label{PWGflux}
G&=&\frac{(2f)^{1/2}}{3^{3/2}}\me^{2\ii\varphi}\cos\vartheta
\Bigg(      \dd\varphi\wedge\dd\vartheta       -\frac{\ii\sin2\vartheta}{3-\cos2\vartheta}\dd\varphi\wedge \sigma_3
 \nn
&&-\frac{4\cos^2\vartheta}{(3-\cos2\vartheta)^2}\dd\vartheta\wedge \sigma_3\Bigg)\wedge(\sigma_2-\ii\sigma_1)~.
\eea
We introduce the Euler angles $(\alpha,\beta,\gamma)$ on $SU(2)$ (as in \cite{Gauntlett:2005ww}),
so that
\bea
\sigma_1 &=& -\sin\gamma\diff\alpha-\cos\gamma\sin\alpha\diff\beta~,\nn
\sigma_2 &=& \cos\gamma\diff\alpha -\sin\gamma\sin\alpha\diff\beta~,\nn
\sigma_3 &=& \diff\gamma-\cos\alpha\dd \beta~.
\eea
In terms of these coordinates, the $R$-symmetry vector $\xi_v$ is \cite{Gauntlett:2005ww}
\bea\label{reeby}
\xi_v=\frac{3}{2}\partial_\varphi - 3\partial_\gamma~.\eea
Using the explicit formulae in \cite{Gauntlett:2005ww}, it is easy to show
that the contact form is
\bea
\sigma = -\frac{2}{3}\left(\cos2\vartheta\, \dd\varphi + \cos^2\vartheta\, \sigma_3\right)~.
\eea

The solution is toric, in the sense that both $\sigma$ and the metric are invariant under
shifts of $\varphi$, $\beta$ and $\gamma$. 
However, notice that the $G$-flux in (\ref{PWGflux}) is not 
invariant under shifts of $\varphi$, thus breaking 
this $U(1)^3$ symmetry to only a $U(1)^2$ symmetry 
of the full supergravity solution. This is expected, since 
the dual field theory described above has only an $SU(2)\times U(1)_R$ 
global symmetry.

On $Y=S^5$ there are precisely three invariant circles under the $U(1)^3$ action, where two
of the $U(1)$ actions degenerate, namely at $\{\vartheta=\tfrac{\pi}{2}\}$,
$\{\vartheta=0,\alpha=0\}$, $\{\vartheta=0,\alpha=\pi\}$.
A set of $2\pi$-period coordinates on $U(1)^3$ are
\bea\label{angulate}
\varphi_1 = \varphi, \quad \varphi_2= -\frac{1}{2}(\varphi+\gamma -\beta), \quad
\varphi_3 = -\frac{1}{2}(\varphi+\gamma+\beta)~.
%\nonumber
\eea
These restrict to coordinates on the
above three invariant circles, respectively.
On $X\cong {\mathbb R}^6\setminus 0$ we also have
three corresponding moment maps
\bea
&& \mu_1=\frac{r^2}{3}\sin^2\vartheta,\quad \mu_2 = \frac{r^2}{3}\cos^2\vartheta
(1+\cos\alpha), \quad\mu_3= \frac{r^2}{3}\cos^2\vartheta(1-\cos\alpha),
%\nonumber
\eea
so that $\omega = \frac{1}{2}\dd(r^2\sigma) = \sum_{i=1}^3 \dd \mu_i\wedge \dd \varphi_i$.
It follows that the image of the moment map -- the space spanned by the
$\mu_i$ coordinates -- is the cone $({\mathbb R}_{\geq 0})^3$, where
the three invariant circles map to the three generating rays
$u_1=(1,0,0)$, $u_2=(0,1,0)$, $u_3=(0,0,1)$.
The Reeb
vector (\ref{reeby}) in this basis is then computed to be
\bea\label{PWReeb}
\xi=\frac{3}{2}\partial_{\varphi}-3\partial_\gamma = \frac{3}{2}\de_{\varphi_1}+\frac{3}{4}\de_{\varphi_2}+\frac{3}{4}\de_{\varphi_3}~.\eea
Since the symplectic structure is smooth at $r=0$, we may evaluate (\ref{DHformula}) by localization without having
to resolve $X$ at $r=0$. In the case at hand, we have the single
fixed point at $r=0$, and from (\ref{PWReeb}) one obtains
the known result
\bea
\frac{a_{{\cal N}=4}}{a_{\mathrm PW}}  = \frac{1}{\xi_1\xi_2\xi_3} = \frac{32}{27}~.
%\nonumber
\eea
They key point about the above calculation is that we have computed
this knowing only the symplectic structure of the solution and the
Reeb vector field $\xi_v$.

We may similarly compute the conformal dimensions
of the operators $\det Z_a$, using (\ref{conformal}),
by interpreting them as arising from a BPS D3-brane wrapped on
the three-spheres at $\alpha=0$ and $\alpha=\pi$, respectively. It is simple
to check these indeed satisfy the calibration condition (\ref{calDBI}) and are thus supersymmetric. Using (\ref{3to4})
and localization at $r=0$ implies that (\ref{3to4}) is equal to
$1/\xi_1\xi_2$, $1/\xi_1\xi_3$, respectively, which in both cases
is $8/9$. The formula (\ref{conformal}) thus gives
$\Delta(\det Z_a)=3N/4$, or equivalently
$\Delta(Z_a)=3/4$, which is indeed the correct result.

Next recall that the complex one-form $\theta=\diff(r^3\theta_0)$, where
$\theta_0 = -\frac{1}{24}\me^{4\Delta} S$, and 
the mesonic moduli space should be the locus $\theta=0$. 
As discussed in section \ref{sec:type}, this is the 
locus $\sin2\Btheta=\sin2\Bphi=0$. 
For the Pilch-Warner solution, we may easily compute
\bea
\sin2\Btheta =
-\frac{\sqrt{3}\sin^2\vartheta}{\sqrt{1+3\sin^4\vartheta}}~,\quad
\cos2\Bphi = \frac{\sqrt{1+3\sin^4\vartheta}}{1+\sin^2\vartheta}~.
\eea
Thus, as discussed in~\cite{Minasian:2006hv}, the mesonic moduli space
$S=0$ is equivalent to $\vartheta=0$, which is a codimension two
submanifold in $\R^6$ diffeomorphic to $\R^4$. Moreover, this is
$\C^2$ in the induced complex structure, and we thus see explicit
agreement with the field theory $N=1$ mesonic moduli space.

Finally, although the Pilch-Warner solution
is generalized complex, rather than complex, we note that
one can nevertheless define a natural complex structure \cite{Halmagyi:2004jy}.
The relation between this integrable complex structure and the
generalized geometry has been discussed in \cite{Minasian:2006hv}. 
Let us conclude this section by elucidating this connection. One can
introduce the following complex coordinates \cite{Minasian:2006hv} in
terms of the angular variables (\ref{angulate}): 
\bea\label{PWs1s2s3}
s_1&=&r^{3/2}\sin\vartheta\, \me^{-\ii\varphi_1}~,\nn
s_2&=&r^{3/4}\cos\vartheta \cos \tfrac{\alpha}{2}\, \me^{\ii\varphi_2}~,\nn
s_3&=&r^{3/4}\cos\vartheta \sin \tfrac{\alpha}{2}\, \me^{\ii\varphi_3}~.
\eea
This makes $\R^6\cong\C^3$. However, because of the minus sign
in the first coordinate in (\ref{PWs1s2s3}),
the corresponding integrable complex structure $I_*$ is \emph{not} the unique complex
structure that is compatible with the toric structure of the solution: the latter instead has complex coordinates
$\bar{s}_1,{s}_2,{s}_3$. Indeed, also the Reeb vector field
$\xi_v$ is \emph{not} given by $I_*(r\de_r)$. This makes the physical
significance of this complex structure rather unclear.
Nevertheless, one can show that $I_*$ does
in fact come from an $SU(3)$ structure defined by a
Killing spinor. Following \cite{Minasian:2006hv}, we define
\bea
   2\hat{a}\eta_* =\eta_+^1 + \ii \eta_+^2
      &= \me^{A/2} \begin{pmatrix}
         \xi_2 \\ \ii\xi_2\end{pmatrix}~,
\eea
where by definition we require $\bar \eta_*\eta_*=1$. It is then
convenient to define $\hat a\equiv |\hat a|\me^{\ii z}$,
where $|\hat{a}|^2 = \frac{1}{2}\me^A |\xi_2|^2 = \frac{1}{2}\me^A (1-\sin\zeta)$. We then introduce the bilinears corresponding to the $SU(3)$ structure defined by $\eta_*$:
\bea
J_* &\equiv &-\ii \bar \eta_* \gamma_{(2)}\eta_*~,\\
\Omega_* &\equiv & \bar\eta^c_*\gamma_{(3)}\eta_*~.
\eea
One computes that $\diff\Omega_*=0$, implying that the corresponding complex structure $I_*$ is integrable, and moreover that
\bea
\me^{2\ii z}\Omega_*=
-\me^{2\ii\alpha}\frac{{\sqrt 2}f^{3/2}}{9\me^{3A}}\dd s^1 \wedge\dd s^2 \wedge\dd s^3~,
\eea
implying that (\ref{PWs1s2s3}) are indeed complex coordinates
for this complex structure. We also compute
\bea
J_*&=&-\frac{\me^{2A}}{r^2}\Big[\diff\log r \wedge \frac{2}{3}\left(\diff\varphi + \frac{\cos^2\vartheta}{(1+\sin^2\vartheta)}{\sigma}_3\right)
+ \frac{1}{3(1+\sin^2\vartheta)}\big(\sin2\vartheta {\sigma}_3\wedge\diff\vartheta \nonumber \\
&& + \cos^2\vartheta{\sigma}_1\wedge{\sigma}_2\big)\Big]~.
\eea

%%%%%%%%%%%%%%%%%%%%%%%%%%%%%%%%%%%%%%%%%%%%%%%%%%%%%%%%%%%%%%%%%%%%%%%%%%%%%%%
\section{Conclusion}
%%%%%%%%%%%%%%%%%%%%%%%%%%%%%%%%%%%%%%%%%%%%%%%%%%%%%%%%%%%%%%%%%%%%%%%%%%%%%%%
In this paper we have initiated an analysis of the generalized cone geometry associated 
with supersymmetric $AdS_5\times Y$ solutions of type IIB supergravity. The cone
is generalized Hermitian and generalized Calabi-Yau and we have identified holomorphic generalized vector fields
that are dual to the dilatation and $R$-symmetry of the dual SCFT. 
We identified a relationship between ``BPS polyforms'', {\it i.e.} polyforms with equal $R$-charge and scaling 
weight, and generalized holomorphic polyforms that should be worth exploring further. In particular, we
would like to make a precise connection between such objects and the 
the spectrum of chiral operators in the SCFT via Kaluza-Klein reduction on $Y$. 

We also showed how one can carry out a generalized reduction of the six-dimensional cone to obtain a new 
four-dimensional transverse generalized Hermitian geometry. This generalizes the transverse K\"ahler-Einstein 
geometry in the Sasaki-Einstein case. By analogy with the Sasaki-Einstein case ({\it e.g.} \cite{Gauntlett:2004hh}) this perspective 
could be useful for constructing new explicit solutions. 

We also analysed the symplectic structure on the cone geometry, which exists providing that
the five-form flux is non-vanishing. It would be interesting to know whether or not this includes all solutions. 
We obtained Duistermaat-Heckman type integrals for the central charge
of the dual SCFT and the conformal dimensions of operators dual to BPS wrapped D3-branes.
These formulae precisely generalize analogous formulae that were derived in \cite{Martelli:2005tp, Martelli:2006yb} for the Sasaki-Einstein case.
Other formulae for these quantities were also presented in \cite{Martelli:2005tp, Martelli:2006yb}  and we expect that these will also 
have precise generalizations in terms of generalized geometry. In particular, we expect a generalized geometric interpretation of
$a$-maximization. 

\subsection*{Acknowledgments}
\noindent
We would like to thank Nick Halmagyi for a useful discussion. M.G. is supported by the Berrow Foundation, J.P.G. by an EPSRC
Senior Fellowship and a Royal Society Wolfson Award, E.P. by a STFC Postdoctoral Fellowship and
J.F.S. by a Royal Society University Research Fellowship.

\appendix
\section{Conventions and 6D to 5D map}\label{sec:56}
We use exactly\footnote{Although it will not be relevant in this paper
  we point out that there is a typo in \cite{Gauntlett:2005ww}: the
  $\rho_a$ matrices generating $\Cliff(4,1)$ actually satisfy
  $\rho_{01234}=-\ii$.} 
the same conventions as in \cite{Gauntlett:2005ww}, up to some simple
relabelling. Here we will explain how the results of that paper
concerning the five-dimensional geometry with metric $g_Y$ can be
uplifted to six-dimensions. In particular, we will relate the
five-dimensional Killing spinors discussed in \cite{Gauntlett:2005ww}
to the six-dimensional chiral spinors $\eta^i$ that define the
bispinors $\Phi_\pm$. We first recall the Killing spinor equations in
five-dimensions, related to the geometry $g_Y$, given in
\cite{Gauntlett:2005ww}. There are two differential conditions 
\begin{align}
   (\nabla_m-\frac{\ii}{2}Q_m) \xi_1
      + \frac{\ii}{4} \left(\me^{-4\Delta}f-2\right)\beta_m \xi_1
      + \frac{1}{8} \me^{-2\Delta} G_{mnp}\beta^{np}\xi_2
      &= 0~, \label{sone}\\
   (\nabla_m+\frac{\ii}{2}Q_m) \xi_2
      -  \frac{\ii}{4} \left(\me^{-4\Delta}f+2\right)\beta_m \xi_2
      + \frac{1}{8} \me^{-2\Delta} G_{mnp}^*\beta^{np}\xi_1
      &= 0~,\label{stwo}
\end{align}
and four algebraic conditions
\begin{align}
   \beta^m\de_m\Delta\xi_1
      - \frac{1}{48}\me^{-2\Delta}\beta^{mnp}G_{mnp}\xi_2
      - \frac{\ii}{4}\left(\me^{-4\Delta}f-4\right) \xi_1
      &= 0 ~, \label{sthree}\\
   \beta^m\de_m\Delta\xi_2
      - \frac{1}{48}\me^{-2\Delta}\beta^{mnp}G_{mnp}^*\xi_1
      + \frac{\ii}{4}\left(\me^{-4\Delta}f+4\right)\xi_2
      &= 0~, \label{sfour}\\
   \beta^m P_m \xi_2
      + \frac{1}{24} \me^{-2\Delta} \beta^{mnp} G_{mnp} \xi_1
      &= 0~, \label{sfive}\\
   \beta^m P_m^* \xi_1
      + \frac{1}{24} \me^{-2\Delta} \beta^{mnp} G_{mnp}^* \xi_2
      &= 0 \label{ssix}~.
\end{align}
Here\footnote{Notice we have relabelled $\gamma_i\mapsto \beta_m$ in \cite{Gauntlett:2005ww}, as in this paper we want to keep the notation $\gamma_i$ for six-dimensional gamma matrices.} the $\beta_m$ generate the Clifford algebra for $g_Y$, so $\{\beta_m,\beta_n\}=2g_{Ymn}$.
Equivalently, with respect to any orthonormal frame the corresponding
$\hat{\beta}_m$ satisfy
$\{\hat{\beta}_m,\hat{\beta}_n\}=2\delta_{mn}$. 
We have chosen $\hat{\beta}_{12345}=+1$. In addition we have set the parameter $m$
in \cite{Gauntlett:2005ww} to be $m=1$, consistent with \reef{normads}.
In the usual string theory variables we have
\bea
\label{psandqs}
P&=&\frac{1}{2}\dd\phi+\frac{\ii}{2}\me^{\phi}F_1~,\nn
Q&=&-\frac{1}{2} \me^\phi F_1~,\nn
G&=& -\ii \me^{\phi/2}F_3-\me^{-\phi/2}H~,
\eea
where the RR field strengths $F_n$ are defined by (\ref{RRforms}).
We also note that the constant $f$ appearing in the Killing spinor equations
is related to the component of the self-dual five-form flux on $Y$
(\ref{5flux}) via 
\be
F_5|_Y=-f\widetilde{\vol}_Y~, 
\ee
where the five-dimensional volume form is defined as
$\widetilde{\vol}_Y=-e^{12345}$ and $e^i$ is the orthonormal frame introduced in
appendix B of \cite{Gauntlett:2005ww}. 

We now provide a map between the five-dimensional spinors and Killing
spinor equations \reef{sone}-\reef{ssix} to six-dimensional
quantities. We begin by using the $\Cliff(5)$ gamma matrices
$\hat{\beta}_{m}$ to construct $\Cliff(6)$ gamma matrices
$\hat{\gamma}_{i}$, $i=1,\dots,6$, via 
\bea
\hat{\gamma}_{m}&=&\hat{\beta}_{m}\otimes \sigma_3~,\qquad m=1,\ldots,5\nn
\hat{\gamma}_6&=&1\otimes\sigma_1~,
\eea
where $\sigma_\alpha$, $\alpha=1,2,3$, are the Pauli matrices.
These satisfy $\{\hat{\gamma}_i,\hat{\gamma}_{j}\}=2\delta_{ij}$.
The corresponding gamma matrices for the six-dimensional
metric $g_6$ will be denoted $\gamma_i$.
We define the 6D chirality operator to be
\be
\tilde\gamma \equiv -\ii\hat{\gamma}_{123456}=1\otimes\sigma_2~.
\ee
We may choose the $D_6$ intertwiner
\be
D_6=\tilde D_5\otimes \sigma_2~,
\ee
where $\tilde D_5=D_5=C_5$ is the intertwiner of $\Cliff(5)$ discussed in
\cite{Gauntlett:2005ww}, and one checks $D_6^{-1}\gamma_i D_6=-\gamma_i^{*}$.
We also note that since in \cite{Gauntlett:2005ww} the interwiner $A_5=1$
we have $A_6=1$ and $\gamma_i^{\dagger}=\gamma_i$.
If $\eta_+$ is a Weyl spinor, satisfying $\tilde\gamma\eta_+ = \eta_+$, then
the conjugate spinor $\eta_-\equiv\eta_+^c\equiv D_6\eta_+^{*}$ satisfies $\tilde\gamma\eta_- = -\eta_-$.

To construct the relevant 6D spinors we first write
\be
\xi_1=\chi_1+\ii\chi_2,\qquad \xi_2=\chi_1-\ii\chi_2~,
\ee
as in \cite{Gauntlett:2005ww}. Given this, the normalization for $\xi_i$ chosen in
\cite{Gauntlett:2005ww} implies that the $\chi_i$ are normalized as
\begin{equation}\label{norminold}
   \bar{\chi}_1\chi_1 = \bar{\chi}_2\chi_2 =
      \tfrac{1}{2}~.
\end{equation}
We then define
\bea\label{Aaronansatz}
\eta_+^1 & = & \me^{A/2} \left(\begin{array}{c} \chi_1 \\ \ii\chi_1 \end{array}\right),
\qquad
\eta_-^1  =  \me^{A/2} \left(\begin{array}{c} -\chi_1^{c} \\ \ii\chi_1^{c} \end{array}\right),
\nonumber\\
\eta_+^2 & = & \me^{A/2} \left(\begin{array}{c} -\chi_2 \\ -\ii\chi_2\end{array}\right),
\qquad
\eta_-^2  =  \me^{A/2} \left(\begin{array}{c} \chi_2^{c} \\ -\ii\chi_2^{c} \end{array}\right)~,
\eea
where recall from (\ref{warp}) that
\bea
\me^{A/2}  = r^{1/2} \me^{\Delta/2+\phi/8}~,
\eea
and also from \cite{Gauntlett:2005ww} that
\bea
\chi^c \equiv \tilde{D}_5 \chi^*~.
\eea
In the conventions of \cite{Gauntlett:2005ww} we have $\bar\chi=\chi^\dagger$.

After some detailed calculation one finds that the five-dimensional Killing spinor equations \reef{sone}--\reef{ssix},
using the five-dimensional metric $g_Y$, are equivalent to the
six-dimensional Killing spinor equations, using the six-dimensional
metric $g_6$ in \reef{6metric} and volume form~\eqref{6vol},
given by
\begin{equation}
  \label{eq:intgravB}
  \left(D_i-\frac14 H_i\right) \eta^1_+ +
\frac{\me^\phi}8 \sla F \gamma_i \eta^2_+=0~,
\end{equation}
\begin{equation}
  \label{eq:extgravB}
\frac12 \,\me^{A}\, \sla\del A\, \eta_+^1 -\frac18 \me^{A+\phi} \sla F\eta^2_+=0~,
\end{equation}
\begin{equation}
  \label{eq:moddilB}
\sla D\eta^1_+ +
\Big(\sla\del (2 A - \phi) -\frac14 \sla H \Big)\eta^1_+=0~,
\end{equation}
and additional equations obtained by applying the rule:
\be\label{swap}
\eta^1 \leftrightarrow \eta^2 \ ,  \quad 
\sla F \to -\sla F^\dagger \ , \quad
H\to -H \ .
\ee
In these equations we are using the notation that, {\it e.g.}
\be
H_i=\tfrac{1}{2}H_{ijk}\gamma^{jk},\qquad \sla F =F_1{}_i\gamma^i+\tfrac{1}{3!}F_3{}_{ijk}\gamma^{ijk}+\tfrac{1}{5!}F_5{}_{ijklm}\gamma^{ijklm}~.
\ee
These are precisely the same equations that were used in \cite{Grana:2006kf} (for zero four-dimensional cosmological constant).

Finally, we record the following equation of \cite{Gauntlett:2005ww}:
\bea\label{ddub}
D(\me^{6\Delta}W) &=& - \me^{6\Delta}P \wedge {W}^* + \frac{f}{4}G\label{a}~,
\eea
where $W$ is the two-form bilinear defined in (\ref{bilins2}).
Using this one can show that
\be
i_{K_5^\#}\left(\frac{4}{f} \me^{6\Delta+\phi/2} \re{W}\right)=  \me^{2\Delta+\tfrac{\phi}{2}}\mathrm{Re}\, K_3~,
\ee
and furthermore that
\be\label{ivh}
\dd \left( \me^{2\Delta+\tfrac{\phi}{2}}\mathrm{Re}\, K_3 \right)=  i_{\xi_v}H~.
\ee
To see the latter one can derive an expression for the left hand side using, amongst other things, (3.18), (3.38) and (B.10) of
\cite{Gauntlett:2005ww}, and an expression for the right hand side using equation (3.38) and (B.8) of \cite{Gauntlett:2005ww}.
Using these results we can deduce that 
\bea\label{lds}
{\cal L}_{K_5^\#}B &=& \dd(i_{K_5^\#}b_2)~, \nn
{\cal L}_{K_5^\#}C_2 &=& \dd(i_{K_5^\#}c_2)~,
\eea
where $b_2$, $c_2$ were introduced in (\ref{gaugeb}), (\ref{gaugec}),
respectively.
%%%%%%%%%%%%%%%%%%%%%%%%%%%%%%%%%%%%%%%%%%%%%%%%%%%%%%%%%%%%
\section{More on the generalized vectors $\xi$ and $\eta$}\label{sec:Rsymappendix}
%%%%%%%%%%%%%%%%%%%%%%%%%%%%%%%%%%%%%%%%%%%%%%%%%%%%%%%%%%%%
In this appendix we derive an expression for the generalized vector $\xi$ in terms of the bilinears
introduced in \cite{Gauntlett:2005ww}. We also use the results of
\cite{Gauntlett:2005ww} to show that $\Lgen_\xi\mathcal{J}_\pm=0$.

The projections of $\xi$ onto the vector and form parts (in a fixed trivialization of $E$)
are denoted $\xi_v$, $\xi_f$, respectively. It will also be convenient to
introduce $\xi^B\equiv \me^B\xi$ whose form part is given by
\be\label{blin}
\xi^{B}_f=\xi_f-i_{\xi_v}B~.
\ee
and we recall that $\xi^B_v=\xi_v$.
We next construct the following two generalized $(1,0)_-$ vectors, which, by definition, are
in the $+\ii$ eigenspace of $\mathcal{J}_-$:
\bea
Z_1^- &=& r\del_r - \ii \xi~, \nn
Z_2^- &=& \dd \log r - \ii \eta~.\label{annomm}
\eea
That is, both are in the annihilator of $\Omega_-$.
We may similarly also construct the $(1,0)_+$ vectors, with respect to $\mathcal{J}_+$:
\bea
Z_1^+ &=& \me^{-\Delta- \phi/4 } r\de_r - \ii \me^{\Delta+ \phi/4 } \eta~, \nn
Z_2^+ &=& \me^{\Delta+ \phi/4 } \dd \log r - \ii \me^{-\Delta- \phi/4 }\xi~. \label{annomp}
\eea
Together $Z^{\pm}_i$ are four independent generalized vectors.
We next note that since $Z^\pm_i$
live within null isotropic subspaces we have six relations of the form, using the notation of (\ref{eq:eta}),
\be
\left<Z_i^{\pm} , Z_j^{\pm}\right> = 0 \ . 
\ee
Explicitly we have 
\bea\label{isoconst}
 i_{\xi_v} \xi_f &=& 0 \;, \qquad
 i_{\xi_v} \dd\log r = 0 \;, \qquad
 i_{r\del_r} \xi_f = 0 \;,\nonumber\\
 i_{\eta_v} \eta_f &=& 0 \;, \qquad
 i_{\eta_v} \dd\log r = 0 \;, \qquad
 i_{r\del_r} \eta_f = 0 \;, \qquad
\left< \xi, \eta  \right> = \tfrac{1}{2} \;.
\eea

Since $Z_1^-$ annihilates $\Omega_-$, using the definition \reef{relomphi} we deduce that
\begin{equation}
\label{annih-rel}
   i_{r\del_r}\Phi_-
      = \ii\left( i_{\xi_{v}}\Phi_- + \xi^{B}_f\wedge\Phi_- \right)~.
\end{equation}
To proceed we use (\ref{Aaron}) to write
\begin{equation}
   \Phi_- \equiv \eta_+^1\otimes\bar\eta_-^{2}
      = \me^A \chi_1\bar{\chi}_2^c \otimes (\sigma_3+\ii\sigma_1)\ .
\end{equation}
Since  $\sla\Phi_-=\sum_{\text{odd $n$}}\tfrac{1}{n!}\Phi_{i_1\dots i_n}\gamma^{i_1\dots
  i_n}$ we have
\begin{equation}
   i_v \Phi_- = \tfrac{1}{2}\{v^i\gamma_i, \Phi_- \}~, \qquad
   \omega\wedge\Phi_- = \tfrac{1}{2}[\omega_i \gamma^i, \Phi_- ]~.
\end{equation}
Hence, using the Clifford algebra decomposition~\eqref{6to5} and
metric \reef{6metric} we have
\begin{equation}
   i_{r\del_r}\Phi_-
       = \tfrac{1}{2}\{ \me^{\Delta+ \phi/4 } \hat{\gamma}_6, \Phi_- \}
       = \tfrac{1}{2} \me^{A+\Delta+ \phi/4 }\chi_1\bar{\chi}_2^c \otimes
          \{ \sigma_1, \sigma_3+\ii\sigma_1\}
       = \ii \me^{A+\Delta+ \phi/4 }\chi_1\bar{\chi}_2^c \otimes 1 \ .
\end{equation}
On the other hand using \reef{isoconst}
we have
\begin{equation}
\begin{aligned}
i_{\xi_{v}}\Phi_- + \xi^{B}_f\wedge\Phi_-
      &= \tfrac{1}{2}\{ \me^{\Delta+ \phi/4 } \xi_v^m\beta_m\otimes\sigma_3, \Phi_- \}
         +\tfrac{1}{2}[ \me^{-\Delta- \phi/4 } \xi_{fm}^{B}\beta^m\otimes\sigma_3, \Phi_- ] \\
      &= \me^{\Delta+ \phi/4 } v^+_m \beta^m\otimes\sigma_3 \Phi_-
         + \me^{\Delta+ \phi/4 } v^-_m \Phi_- \beta^m\otimes\sigma_3 \\
      &= \me^{A+\Delta+ \phi/4 } \left(
           v^+_m \beta^m(\chi_1\bar{\chi}_2^c)
           + v^-_m (\chi_1\bar{\chi}_2^c) \beta^m \right) \otimes 1
        \\ & \qquad \qquad \qquad
        - \me^{A+\Delta+ \phi/4 }\left(
           v^+_m \beta^m(\chi_1\bar{\chi}_2^c)
           - v^-_m (\chi_1\bar{\chi}_2^c) \beta^m \right) \otimes
           \sigma_2 \ ,
\end{aligned}
\end{equation}
where recall that $\{\beta_m,\beta_n\}=2g_{Ymn}$ and we have defined
\bea
v^\pm_m=\tfrac{1}{2}(\xi_{vm}\pm \me^{-2\Delta-\phi/2}\xi^{B}_{fm})~.\eea
To satisfy~\eqref{annih-rel} we thus require
\begin{equation}
   v^+_m \beta^m(\chi_1\bar{\chi}_2^c)
      = v^-_m (\chi_1\bar{\chi}_2^c) \beta^m
      = \tfrac{1}{2}\chi_1\bar{\chi}_2^c~,
\end{equation}
which implies
\begin{equation}
   v^+_m \beta^m \chi_1 = \tfrac{1}{2}\chi_1~, \qquad
   v^-_m \beta^m \chi_2 = \tfrac{1}{2}\chi_2~,
\end{equation}
or equivalently
\begin{equation}
   v^+_m
     = \frac{\bar{\chi}_1\beta_m \chi_1}{2\bar{\chi}_1\chi_1}~,
  \qquad
   v^-_m
     = \frac{\bar{\chi}_2\beta_m \chi_2}{2\bar{\chi}_2\chi_2}~.
\end{equation}
Hence, given the normalizations \reef{norminold} we deduce that, 
in terms of the bilinears defined in \eqref{bilins},
\begin{equation}
\begin{aligned}\label{identofgv}
   \xi_{v} &=K_5^{\#}~,\\
   \xi^{B}_{f} &= \me^{2\Delta+\phi/2}\mathrm{Re}\, K_3 \ .
\end{aligned}
\end{equation}

A similar calculation using
\begin{equation}
   i_{r\del_r}\Phi_+
      = \ii \me^{2\Delta+\phi/2}\left( i_{\eta_{v}}\Phi_- + \eta^{B}_f\wedge\Phi_+ \right)~,
\end{equation}
leads to
\bea
\eta_v &=& \me^{-2\Delta - \phi/2}\re{K_3^{\#}}~,\nn
\eta_f^{B} &=& K_5~.
\eea
Using the expression for the $B$-field given in \reef{gaugeb}
we obtain the expressions for $\xi_f$ and $\eta_f$ given in \reef{tranresvects}.

In \cite{Gauntlett:2005ww} it was shown that $K_5$ is a Killing one-form, so that
its dual vector field $K_5^\#$, with respect to the metric $g_Y$ on $Y$, is a Killing vector field.
In fact $K_5^\#$ generates a full symmetry of the supergravity solution,
in that all bosonic fields (warp factor, dilaton, NS three-form $H$ and RR fields)
are preserved under the Lie derivative along $\xi_v=K_5^\#$.
However, importantly, the Killing spinors $\xi_1$, $\xi_2$ are not
invariant under $\xi_v$. In \cite{Gauntlett:2005ww} it was shown that
\bea\label{Scharge}
\mathcal{L}_{\xi_v} S =  -3\ii S~,
\eea
where $S \equiv  \bar{\xi}_2^c \xi_1$.
Notice that, since $\xi_v$ preserves all of the bosonic fields,
one may take the Lie derivative of the Killing spinor
equations (\ref{sone})-(\ref{ssix}) for $\xi_1$, $\xi_2$ along $\xi_v$, showing
that $\{\mathcal{L}_{\xi_v} \xi_i\}$ satisfy the same
equations as the $\{\xi_i\}$.
It thus follows that
\bea
\mathcal{L}_{\xi_v} \xi_i  =  \ii\mu  \xi_i~,\eea
where $\mu$ is a constant. Now (\ref{Scharge}) implies that
$2\mu = -3$, and thus
\bea
\mathcal{L}_{\xi_v}  \xi_i =  -\frac{3\ii}{2} \xi_i~.
\eea
One can also derive this last equation directly from the Killing spinor
equations (\ref{sone})-(\ref{ssix}) of \cite{Gauntlett:2005ww}.
It thus follows that
\bea
\mathcal{L}_{\xi_v} \, \Phi_+ & = & 0~,\\
\mathcal{L}_{\xi_v} \, \Phi_- & = & -3\ii\, \Phi_-~.
\eea
From \reef{ivh} we have $\dd\xi^B_f=i_{\xi_v} H$ and we deduce that
\bea
\Lgen_{\xi^B}\Phi_+&=&i_{\xi_v}H\wedge\Phi_+~,\nn
\Lgen_{\xi^B}\Phi_-&=&-3\ii\, \Phi_- +i_{\xi_v}H\wedge\Phi_-~.
\eea
Since \reef{ivh} is also equivalent to $\dd \xi_f =  \mathcal{L}_{\xi_v} B$
we deduce that
\bea
\Lgen_\xi\Omega_+&=&0~,\nn
\Lgen_\xi\Omega_-&=&-3\ii \Omega_-~,
\eea
and hence $\Lgen_\xi {\mathcal J}_\pm=0$.
It is also interesting to point out that
\bea
(\Lgen_{\xi^B} -i_{\xi_v}H\wedge) F=0~, \quad \mathrm{or}\ \, \mathrm{equivalently}~,\quad 
\Lgen_\xi(\me^{-B}F)=0~.
\eea

\section{The Sasaki-Einstein case}
Here we discuss the special  case in which the compact five-manifold $Y$ is Sasaki-Einstein. Setting $G=P=Q=0$, $f=4e^{4\Delta}$ and $\xi_2=0$,   
the Killing spinor equations \reef{sone}-\reef{ssix} reduce to
\be
\nabla_m\xi_1+\frac{\ii}{2}\beta_m\xi_1=0~.
\ee
In terms of appendix B of \cite{Gauntlett:2005ww} we choose $\bar\theta=\bar\phi=0$
and $\me^{2\ii\bar\alpha}=-1$ (these angles had no bars on them in \cite{Gauntlett:2005ww}). 
We then have the equalities
\bea
\eta&=&\frac{1}{2}\bar\xi_1\beta_{(1)}\xi_1=K_5=e^1~,\nn
\omega_{\mathrm{KE}}&=&\frac{\ii}{2}\bar\xi_1\beta_{(2)}\xi_1=-V=e^{25}+e^{43}~,\nn
\Omega_{\mathrm{KE}}&=&\frac{1}{2}\bar\xi_1\beta_{(2)}\xi^c_1=(e^2+\ii e^5)\wedge (e^4+\ii e^3)~,
\eea
and 
\bea
\dd\eta&=&2\omega_{\mathrm{KE}}~,\nn
\dd\Omega_{\mathrm{KE}}&=&3\ii\eta\wedge \Omega_{\mathrm{KE}}~.
\eea
Observe that 
\be
\eta\wedge\frac{1}{2!}\omega^2_{\mathrm{KE}}=-e^{12345}=\widetilde{\vol}_Y~.
\ee

Next using the 5D-6D map \eqref{Aaron}, we obtain
\bea
\ii\bar\eta^1_+\gamma_{(2)}\eta^1_+&=&r(\dd\log r\wedge e^1+\omega_{\mathrm{KE}})\equiv\frac{1}{r}\omega_{\mathrm{CY}}~,\nn
-\ii\bar\eta_+^{1c}\gamma_{(3)}\eta^1_+&=&r(\dd\log r-\ii e^1)(e^2-\ii e^5)(e^4-\ii e^3)\equiv\frac{1}{r^2}\bar\Omega_{\mathrm{CY}}~.
\eea
It is worth noting that
\be
\frac{1}{3!}\omega^3_{\mathrm{CY}}=r^3e^{123456}=r^3\dd\log r\wedge \eta\wedge \frac{1}{2!}\omega_{\mathrm{KE}}^2~.
\ee

We also find, directly from \reef{ps1def}, \reef{Omegaplus}
\bea
\Omega_-&=&\frac{1}{8}\bar\Omega_{\mathrm{CY}}~,\nn
\Omega_+&=&-\frac{\ii r^3}{8}\exp\left(\frac{\ii}{r^2}\omega_{\mathrm{CY}}\right)~.
\eea
A useful check is that these expressions agree with those obtained from
the general expressions obtained in sections~\ref{genO-}
and~\ref{genO+}, respectively. 

One can also write down the corresponding reduced structures $\Lred$
and $\Rred$, as defined in section~\ref{sec:Omega-red}, on the
K\"ahler-Einstein space. One finds
\begin{equation}
   \Lred = \frac{1}{8}\me^{3\ii\psi}\bar{\Omega}_{\mathrm{KE}} \ , 
   \qquad
   \Rred = - \frac{\ii}{8}\me^{\ii\omega_{\mathrm{KE}}} . 
\end{equation}
where $\psi$ is the coordinate, defined such that $K_5^\#=\del_\psi$,
introduced in~\eqref{psi-def}.


\begin{thebibliography}{99}

%\cite{Gauntlett:2004zh}
\bibitem{Gauntlett:2004zh}
  J.~P.~Gauntlett, D.~Martelli, J.~Sparks and D.~Waldram,
  ``Supersymmetric $AdS_5$ solutions of M-theory,''
  Class.\ Quant.\ Grav.\  {\bf 21} (2004) 4335
  [arXiv:hep-th/0402153].
  %%CITATION = CQGRD,21,4335;%%

%\cite{Gauntlett:2004yd}
\bibitem{Gauntlett:2004yd}
  J.~P.~Gauntlett, D.~Martelli, J.~Sparks and D.~Waldram,
  ``Sasaki-Einstein metrics on $S_2\times S_3$,''
  Adv.\ Theor.\ Math.\ Phys.\  {\bf 8} (2004) 711
  [arXiv:hep-th/0403002].
  %%CITATION = 00203,8,711;%%

%\cite{Cvetic:2005ft}
\bibitem{Cvetic:2005ft}
  M.~Cvetic, H.~Lu, D.~N.~Page and C.~N.~Pope,
  ``New Einstein-Sasaki spaces in five and higher dimensions,''
  Phys.\ Rev.\ Lett.\  {\bf 95} (2005) 071101
  [arXiv:hep-th/0504225].
  %%CITATION = PRLTA,95,071101;%%

%\cite{Benvenuti:2004dy}
\bibitem{Benvenuti:2004dy}
  S.~Benvenuti, S.~Franco, A.~Hanany, D.~Martelli and J.~Sparks,
  ``An infinite family of superconformal quiver gauge theories with
  Sasaki-Einstein duals,''
  JHEP {\bf 0506} (2005) 064
  [arXiv:hep-th/0411264].
  %%CITATION = JHEPA,0506,064;%%

%\cite{Hanany:2005ve}
\bibitem{Hanany:2005ve}
  A.~Hanany and K.~D.~Kennaway,
  ``Dimer models and toric diagrams,''
  arXiv:hep-th/0503149.
  %%CITATION = HEP-TH/0503149;%%

%\cite{Franco:2005rj}
\bibitem{Franco:2005rj}
  S.~Franco, A.~Hanany, K.~D.~Kennaway, D.~Vegh and B.~Wecht,
  ``Brane Dimers and Quiver Gauge Theories,''
  JHEP {\bf 0601} (2006) 096
  [arXiv:hep-th/0504110].
  %%CITATION = JHEPA,0601,096;%%

%\cite{Franco:2005sm}
\bibitem{Franco:2005sm}
  S.~Franco, A.~Hanany, D.~Martelli, J.~Sparks, D.~Vegh and B.~Wecht,
  ``Gauge theories from toric geometry and brane tilings,''
  JHEP {\bf 0601} (2006) 128
  [arXiv:hep-th/0505211].
  %%CITATION = JHEPA,0601,128;%%

%\cite{Anselmi:1997ys}
\bibitem{Anselmi:1997ys}
  D.~Anselmi, J.~Erlich, D.~Z.~Freedman and A.~A.~Johansen,
  ``Positivity constraints on anomalies in supersymmetric gauge theories,''
  Phys.\ Rev.\  D {\bf 57}, 7570 (1998)
  [arXiv:hep-th/9711035].
  %%CITATION = PHRVA,D57,7570;%%

\bibitem{kenbrian}  K.~A.~Intriligator and B.~Wecht,
  ``The exact superconformal R-symmetry maximizes $a$,''
  Nucl.\ Phys.\  B {\bf 667}, 183 (2003),
  [arXiv:hep-th/0304128].
  %%CITATION = NUPHA,B667,183;%%

%\cite{Martelli:2005tp}
\bibitem{Martelli:2005tp}
  D.~Martelli, J.~Sparks and S.~T.~Yau,
  ``The geometric dual of $a$-maximisation for toric Sasaki-Einstein
  manifolds,''
  Commun.\ Math.\ Phys.\  {\bf 268}, 39 (2006)
  [arXiv:hep-th/0503183].
  %%CITATION = CMPHA,268,39;%%

%\cite{Martelli:2006yb}
\bibitem{Martelli:2006yb}
  D.~Martelli, J.~Sparks and S.~T.~Yau,
  ``Sasaki-Einstein manifolds and volume minimisation,''
  Commun.\ Math.\ Phys.\  {\bf 280}, 611 (2008)
  [arXiv:hep-th/0603021].
  %%CITATION = CMPHA,280,611;%%

%\cite{Lunin:2005jy}
\bibitem{Lunin:2005jy}
  O.~Lunin and J.~M.~Maldacena,
  ``Deforming field theories with $U(1)\times U(1)$ global symmetry and their
  gravity duals,''
  JHEP {\bf 0505}, 033 (2005)
  [arXiv:hep-th/0502086].
  %%CITATION = JHEPA,0505,033;%%

%\cite{Pilch:2000ej}
\bibitem{Pilch:2000ej}
  K.~Pilch and N.~P.~Warner,
  ``A new supersymmetric compactification of chiral IIB supergravity,''
  Phys.\ Lett.\  B {\bf 487}, 22 (2000)
  [arXiv:hep-th/0002192].
  %%CITATION = PHLTA,B487,22;%%

%\cite{Khavaev:1998fb}
\bibitem{Khavaev:1998fb}
  A.~Khavaev, K.~Pilch and N.~P.~Warner,
  ``New vacua of gauged $N = 8$ supergravity in five dimensions,''
  Phys.\ Lett.\  B {\bf 487} (2000) 14
  [arXiv:hep-th/9812035].
  %%CITATION = PHLTA,B487,14;%%

%\cite{Halmagyi:2005pn}
\bibitem{Halmagyi:2005pn}
  N.~Halmagyi, K.~Pilch, C.~Romelsberger and N.~P.~Warner,
  ``Holographic duals of a family of $N = 1$ fixed points,''
  JHEP {\bf 0608}, 083 (2006)
  [arXiv:hep-th/0506206].
  %%CITATION = JHEPA,0608,083;%%

%\cite{Leigh:1995ep}
\bibitem{Leigh:1995ep}
  R.~G.~Leigh and M.~J.~Strassler,
  ``Exactly Marginal Operators And Duality In Four-Dimensional $N=1$
  Supersymmetric Gauge Theory,''
  Nucl.\ Phys.\  B {\bf 447}, 95 (1995)
  [arXiv:hep-th/9503121].
  %%CITATION = NUPHA,B447,95;%%

%\cite{Benvenuti:2005wi}
\bibitem{Benvenuti:2005wi}
  S.~Benvenuti and A.~Hanany,
  ``Conformal manifolds for the conifold and other toric field theories,''
  JHEP {\bf 0508} (2005) 024
  [arXiv:hep-th/0502043].
  %%CITATION = JHEPA,0508,024;%%

%\cite{Aharony:2002hx}
\bibitem{Aharony:2002hx}
  O.~Aharony, B.~Kol and S.~Yankielowicz,
  ``On exactly marginal deformations of $N = 4$ SYM and type IIB
  supergravity on $AdS_5\times S^5$,''
  JHEP {\bf 0206} (2002) 039
  [arXiv:hep-th/0205090].
  %%CITATION = JHEPA,0206,039;%%

%\cite{Gauntlett:2005ww}
\bibitem{Gauntlett:2005ww}
  J.~P.~Gauntlett, D.~Martelli, J.~Sparks and D.~Waldram,
  ``Supersymmetric $AdS_5$ solutions of type IIB supergravity,''
  Class.\ Quant.\ Grav.\  {\bf 23}, 4693 (2006)
  [arXiv:hep-th/0510125].
  %%CITATION = CQGRD,23,4693;%%

%\cite{Minasian:2006hv}
\bibitem{Minasian:2006hv}
  R.~Minasian, M.~Petrini and A.~Zaffaroni,
  ``Gravity duals to deformed SYM theories and generalized complex geometry,''
  JHEP {\bf 0612}, 055 (2006)
  [arXiv:hep-th/0606257].
  %%CITATION = JHEPA,0612,055;%%

%\cite{Butti:2007aq}
\bibitem{Butti:2007aq}
  A.~Butti, D.~Forcella, L.~Martucci, R.~Minasian, M.~Petrini and A.~Zaffaroni,
  ``On the geometry and the moduli space of beta-deformed quiver gauge
  theories,''
  JHEP {\bf 0807}, 053 (2008)
  [arXiv:0712.1215 [hep-th]].
  %%CITATION = JHEPA,0807,053;%%

\bibitem{Grana:2004bg}
  M.~Grana, R.~Minasian, M.~Petrini and A.~Tomasiello,
  ``Supersymmetric backgrounds from generalized Calabi-Yau manifolds,''
  JHEP {\bf 0408}, 046 (2004) [arXiv: hep-th/0406137].
  %%CITATION = JHEPA,0408,046;%%

%\cite{Grana:2005sn}
\bibitem{Grana:2005sn}
  M.~Grana, R.~Minasian, M.~Petrini and A.~Tomasiello,
  ``Generalized structures of $N=1$ vacua,''
  JHEP {\bf 0511}, 020 (2005) 
  [arXiv:hep-th/0505212].
  %%CITATION = JHEPA,0511,020;%%

\bibitem{hitchin} N.~Hitchin, ``Generalized Calabi-Yau manifolds,''
   Quart. J. Math. Oxford Ser.
   {\bf 54} (2003) 281-308 [arXiv:math/0209099].

%\cite{Gauntlett:2004hh}
\bibitem{Gauntlett:2004hh}
  J.~P.~Gauntlett, D.~Martelli, J.~F.~Sparks and D.~Waldram,
  ``A new infinite class of Sasaki-Einstein manifolds,''
  Adv.\ Theor.\ Math.\ Phys.\  {\bf 8} (2006) 987
  [arXiv:hep-th/0403038].
  %%CITATION = 00203,8,987;%%

\bibitem{bcg1}
  H.~Bursztyn, G.~Cavalcanti and  M.~Gualtieri
  ``Reduction of Courant algebroids and generalized complex
  structures'', 
  Adv.\ Math.\, {\bf 211}, 726 (2007) [arXiv: math/0509640v3 [math.DG]]. 

\bibitem{bcg}
   H.~Bursztyn, G.~Cavalcanti and  M.~Gualtieri
   ``Generalized Kahler and hyper-Kahler quotients'',
   arXiv:math/0702104v1 [math.DG].

%\cite{Gabella:2009ni}
\bibitem{shortpaper}
 M.~Gabella, J.~P.~Gauntlett, E.~Palti, J.~Sparks and D.~Waldram,
  ``The central charge of supersymmetric AdS$_5$ solutions of type IIB
  supergravity,''
  Phys.\ Rev.\ Lett.\  {\bf 103} (2009) 051601
  [arXiv:0906.3686 [hep-th]].
  %%CITATION = PRLTA,103,051601;%%

\bibitem{gualtieri} M.~Gualtieri, ``Generalized complex geometry,''
   arXiv:math/0703298.

\bibitem{Hitchin:2005in}
  N.~Hitchin,
  ``Brackets, forms and invariant functionals,''
  arXiv:math/0508618.
  %%CITATION = MATH/0508618;%%

%\cite{Grana:2008yw}
\bibitem{Grana:2008yw}
  M.~Grana, R.~Minasian, M.~Petrini and D.~Waldram,
  ``T-duality, Generalized Geometry and Non-Geometric Backgrounds,''
  arXiv:0807.4527 [hep-th].
  %%CITATION = ARXIV:0807.4527;%%

\bibitem{cavalcanti}
G.~R. Cavalcanti, ``New aspects of the {$dd^c$}-lemma,''
arXiv:math.dg/0501406.
%%CITATION = MATH.DG/0501406;%%.

\bibitem{gualtierithesis} M.~Gualtieri, ``Generalized complex geometry,'' Oxford University DPhil thesis, 
arXiv:math/0401221 [math-DG].

%\cite{Grana:2006kf}
\bibitem{Grana:2006kf}
  M.~Grana, R.~Minasian, M.~Petrini and A.~Tomasiello,
  ``A scan for new $N=1$ vacua on twisted tori,''
  JHEP {\bf 0705} (2007) 031
  [arXiv:hep-th/0609124].
  %%CITATION = JHEPA,0705,031;%%

%\cite{Martucci:2005ht}
\bibitem{Martucci:2005ht}
  L.~Martucci and P.~Smyth,
  ``Supersymmetric D-branes and calibrations on general N = 1 backgrounds,''
  JHEP {\bf 0511}, 048 (2005)
  [arXiv:hep-th/0507099].
  %%CITATION = JHEPA,0511,048;%%

%\cite{Tomasiello:2007zq}
\bibitem{Tomasiello:2007zq}
  A.~Tomasiello,
  ``Reformulating Supersymmetry with a Generalized Dolbeault Operator,''
  JHEP {\bf 0802}, 010 (2008)
  [arXiv:0704.2613 [hep-th]].
  %%CITATION = JHEPA,0802,010;%%

%\cite{Martucci:2006ij}
\bibitem{Martucci:2006ij}
  L.~Martucci,
  ``D-branes on general $N = 1$ backgrounds: Superpotentials and D-terms,''
  JHEP {\bf 0606}, 033 (2006)
  [arXiv:hep-th/0602129].
  %%CITATION = JHEPA,0606,033;%%

\bibitem{Henningson:1998gx}
  M.~Henningson and K.~Skenderis,
  ``The holographic Weyl anomaly,''
  JHEP {\bf 9807}, 023 (1998)
  [arXiv:hep-th/9806087].
  %%CITATION = JHEPA,9807,023;%%

%\cite{Martelli:2007mk}
\bibitem{Martelli:2007mk}
  D.~Martelli and J.~Sparks,
  ``Baryonic branches and resolutions of Ricci-flat Kahler cones,''
  JHEP {\bf 0804}, 067 (2008)
  [arXiv:0709.2894 [hep-th]].
  %%CITATION = JHEPA,0804,067;%%

%\cite{Klebanov:2007us}
\bibitem{Klebanov:2007us}
  I.~R.~Klebanov and A.~Murugan,
  ``Gauge/Gravity Duality and Warped Resolved Conifold,''
  JHEP {\bf 0703}, 042 (2007)
  [arXiv:hep-th/0701064].
  %%CITATION = JHEPA,0703,042;%%

%\cite{Halmagyi:2004jy}
\bibitem{Halmagyi:2004jy}
  N.~Halmagyi, K.~Pilch, C.~Romelsberger and N.~P.~Warner,
  ``The complex geometry of holographic flows of quiver gauge theories,''
  JHEP {\bf 0609}, 063 (2006)
  [arXiv:hep-th/0406147].
  %%CITATION = JHEPA,0609,063;%%

\end{thebibliography}
\end{document}